\documentclass[11pt,fleqn]{article}

\title{PointQ model of an arterial network:\ calibration and experiments\thanks{Sam Coogan, Gabriel Gomes, Alex A. Kurzhanskiy, Negar Zahedi Mehr and Roberto Horowitz helped clarify  theoretical issues.  Anthony Patire, Yiguang Xuan and Thomas Schreiter generously provided the data for the case study.
We thank them.  This research was supported by the California Department of Transportation, National Science Foundation SBIR Phase II Award 1329477, and TUBITAK-2219 Program.}
 }
\author{Fatma Yildiz Tascikaraoglu, Jennie Lioris,  Ajith Muralidharan, Martin Gouy\\ and Pravin Varaiya }

\date{\today}

\usepackage{graphicx}

\usepackage{amsmath}
\usepackage{amsfonts}
\usepackage{caption}
\usepackage{pdfpages}
\usepackage{appendix}
\usepackage{caption}
\usepackage[authoryear,round]{natbib}
\usepackage{subcaption}

\renewcommand{\L}{{\cal L}}
\newcommand{\Lout}{{\cal L}_{exit}}
\newcommand{\Lin}{{\cal L}_{entry}}
\newcommand{\Lall}{{\cal L}_{all}}
\newcommand{\N}{{\cal N}}
\newcommand{\M}{{\cal M}}
\newcommand{\U}{{\cal U}}

\newcommand{\E}{{\cal L}}
\newcommand{\G}{{\cal G}}
\renewcommand{\H}{{\cal H}}

\newcommand{\g}{\gamma}

\newtheorem{theorem}{Theorem}
\newtheorem{corollary}{Corollary}
\begin{document}
\maketitle

\begin{abstract}
The calibration of a PointQ arterial microsimulation model is formulated as a quadratic programming problem (QP) whose decision variables are  link flows, demands at entry links, and turn movements at intersections, subject to linear constraints imposed by
 flow conservation identities and field measurements of a subset of link flows (counts), demands and turn ratios.  
The quadratic objective function is the deviation of the  decision variables from their measured values.   
The solution to the QP gives estimates  of all unmeasured variables and  thus yields a fully specified simulation model.    Runs of this  simulation model can then be
compared with other field measurements, such as travel times along routes, to judge the reliability of the calibrated  model.
A  section of the Huntington-Colorado arterial near I-210 in Los Angeles comprising 73 links and  16 intersections is used to illustrate the procedure.  Two experiments are conducted with the calibrated model to determine the maximum traffic that can be diverted from the I-210 freeway to the arterial network, with and without permitting changes in the timing plans.  The maximum diversion in both cases is obtained by solving a linear programming problem.  A third experiment compares the delay and travel time using the existing fixed time control and a max pressure control.  The fourth experiment compares two
PointQ models: in the first model the freeway traffic follows a pre-specified route while the background traffic moves according to turn ratios, 
and in the second model turn ratios are modified in a single-commodity model to match the link flows.  The substantial modification of the turn ratios needed  suggests that the use of a single-commodity model as frequently done in CTM models can be misleading.
Often there are too few measurements to identify all the unmeasured link flows.  A simple graphical condition is provided to check whether or not a particular unmeasured link flow can be identified from the measured flows.  Another condition determines the additional measurements needed to identify all unmeasured flows.
If not all flows can be identified, the magnitude of inaccuracy in terms of VMT is estimated as the solution to a linear programming problem.  Lastly, analysis of congestion as the network loading is increased suggests a macroscopic queuing model of the network, which is compared with the macroscopic fundamental diagram.

\end{abstract}

\section{Introduction} \label{sec-intro}
\vspace*{-0.1in}
Calibration of  an arterial network model often requires estimating the parameters of a microsimulation model.   The typical procedure is to simulate various combinations of model parameters and select that combination whose simulation results best match   field measurements.  One shortcoming of this approach is that the number of parameter combinations is huge whereas field data are too sparse to reliably distinguish between different parameter combinations. 

Suppose, following \cite{bpark}, that one elects to tune a 6-dimensional  parameter vector $\theta$ of a stochastic VISSIM model using  field 
data of travel time measurements  over a route.  Stochastic runs of the simulation model generate samples from a random distribution $f(\theta)$ of the travel time. Since $f(\theta)$ is not explicitly known, 
it is impossible to estimate $\theta \in R^6$ with any certainty  even with several thousand field measurements of travel time.  (If each parameter vector takes 8 discrete values, $\theta$ has $6^8$ or 1.7M possible values.)  Consequently, 
trust is eventually placed  on how well the visualization of a small number of  simulation runs confirms  the analyst's or client's beliefs, rather than on any statistical measure of confidence.

The second difficulty is that the parameters $\theta$ of the VISSIM model such as `emergency stop distance' and `lane change distance' selected by \cite{bpark}, cannot
be intuitively or theoretically related to the field data (travel time) so that one has no sound basis for specifying a parametric form for $f(\theta)$ (\cite{bpark} choose a linear function for $f$).

The third difficulty is that the effort that goes into microsimulation calibration detracts from the goal of the simulation exercise which may be to design say a robust controller or to quickly evaluate a large number of tactical responses to incidents or events.  

The approach explored in this paper is  different and more straightforward.  It adopts a  simple microsimulator, called PointQ (\cite{wodes,liorisTRR}), which models the network of intersections as  a queuing network.  Vehicles arrive from outside the network at entry links in a deterministic or  Poisson 
stream.  They take a fixed or random time to travel along a link at the end of which they join a queue.  Vehicles make turns at intersections either in fixed proportions or according to randomly selected O-D routes.  PointQ is a discrete event simulator whose  important events are associated with vehicle movement from a queue at one intersection to a queue at another intersection, and controller feedback actions that determine which phases are actuated.  

The output of PointQ is the record of all the events in the simulation run.  The records are uploaded into a relational database.
Responses to database queries yield network performance measures such as distribution of travel times along routes and vehicle-miles traveled (VMT), as well as  intersection performance measures such as queuing delay, wasted green and progression quality.  A visualization facility allows one to plot the performance measures as well as create an animation of the movement of vehicles through the network.

Three sets of parameters specify a PointQ model: network geometry (link capacity and travel time or speed limit, turn pocket capacity, saturation flow rates for each phase); signal controller configuration; and specification of demand in terms of turn ratios or OD flows.  The network  and demand parameters (typically turn movement counts) are relatively easy to measure, the controller configuration is obtained from the local transportation agency.  

One advantage of PointQ over (say) VISSIM is that the underlying queuing network model provides a mathematical framework to analyze traffic movement (\cite{FTControl}) and to design signal control (\cite{aboudolas09A,MPtrc}), so   theoretical results based on simplified assumptions can be compared with more complex simulations.  The computational advantage of PointQ  is that like in `mesoscopic' models such as Dynasmart (\cite{jayakrishnan1994}), it does not use a driver model. Of course PointQ should not be used to study (say) the effect of driver aggression on safety or the impact of special roadway geometry.   Further, its transparency makes it  easy to spot PointQ model specification errors.  Lastly, by increasing  network loading, one can observe the onset of  network congestion accompanied by a reduction in the exit flows.  This prompts a formulation of  network aggregate behavior as a macroscopic queuing model, which can be compared with the macroscopic fundamental diagram proposed by \cite{mfd}. 

The rest of the paper is organized as follows.  \S \ref{sec-model} describes PointQ's underlying queuing model.   \S\ref{sec-constraints} presents the flow constraints, and \S \ref{sec-calibration} formulates the calibration task as a quadratic programming problem. \S \ref{sec-case}
presents a calibration case study and  \S \ref{sec-exp} describes experiments that motivated the case study.  \S \ref{sec-indetermin} considers the situation in which there are too few measurements to identify all flows in the network.  \S \ref{sec-macro} proposes a macroscopic queuing model.  \S \ref{sec-conc} collects the conclusions.

\section{Model specification} \label{sec-model}
\vspace*{-0.1in}
A PointQ arterial network model is specified by five  data items:
\vspace*{-0.1in}
\begin{enumerate}
\setlength\itemsep{0in} \parskip0pt \parsep0pt
    \item Network graph, link data: link storage capacity in number of vehicles (including turn pockets), link travel time (constant or stochastic),  and possible movements (phases);
    \item Saturation flow rate for each movement at each intersection;
    \item Demand: external flows at each entry link;
    \item Routing: turn ratios for each movement at every signal and/or O-D routes;  
    \item Specification of  controller parameters at each signalized intersection, such as timing plan, offset, and cycle time for a fixed-time controller.
\end{enumerate}
The calibration procedure assumes a network in `steady state'  described using the following notation, which assumes a fixed time control.  
\begin{eqnarray*}
\Lall&=& \mbox{ set of all links, elements $l, m, k $} \\
\L [\Lin, \Lout]&=& \mbox{ set of internal [entry, exit] links}\\
In_l &\subset& \L \cup \Lin, \mbox{ set of links input to $l\in \L \cup \Lout$} \\
Out_l & \subset& \L \cup \Lout, \mbox{ set of links output from $l \in \L \cup \Lin$} \\
f_l &=& \mbox{ flow of vehicles in link $l$, vehicles per period}\\
d_l &=& \mbox{ exogenous flow of vehicles in link $l \in \Lin$, vehicles per period}\\
f(l,m)&=& \mbox{ flow of vehicles making movement $(l,m)$, i.e. leaving $l$ and entering $m$}, \\
&& \mbox{ vehicles per period} \\
r(l,m) &=& \mbox{ fraction of flow $f_l$ leaving  $l$ and entering  $m$ }\\
c(l,m) &=& \mbox{ saturation flow rate of phase $(l,m)$, vehicles per period}\\
g(l,m) &=& \mbox{ actuation time of phase $(l,m)$} \\
\N &=& \mbox{ set of nodes or intersections, elements $n $} \\
I(n) & \subset&\L\cup \Lin, \mbox{ set of links entering $n $} \\
O(n) & \subset&\L \cup \Lout, \mbox{ set of links leaving $n$} \\
T_n &=&  \mbox { cycle time of intersection $n$} \\
D, M, T &=& \mbox{ indexes of measured demands, flows, turn ratios }
\end{eqnarray*}

\section{Flow constraints} \label{sec-constraints}
\begin{figure}
\centering
\includegraphics[width=3in]{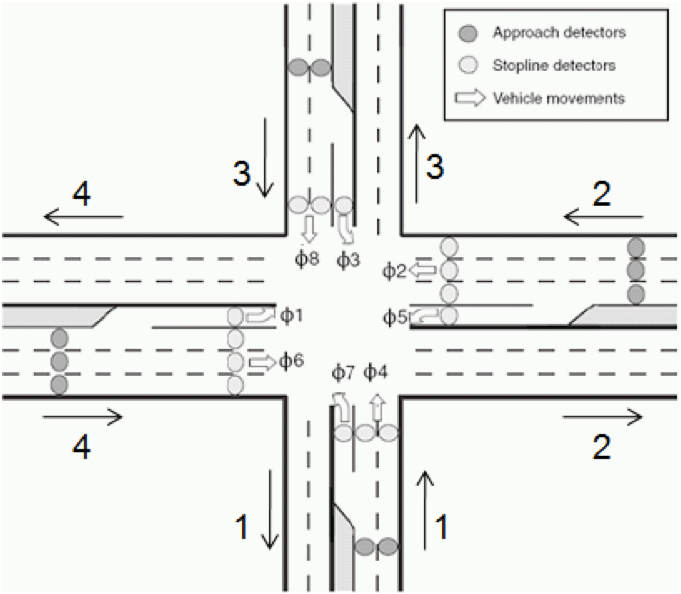}
\caption{Standard intersection with four approaches and four departures.}
\label{fig8}
\end{figure}

\vspace*{-0.1in}
\textbf{Flow conservation }\\
 Demand, routing and flow conservation impose constraints \eqref{1}-\eqref{4}: 
\begin{eqnarray}
f_l &=&d_l, \:l \in \mathcal{L}_{entry} ,\label{1}\\
f(l,m) &=& r(l,m) f_l , \label{2} \\
f_l &=&  \sum_{m}{f(l,m) },  \: l \in  \L \cup \Lin  ,\label{3} \\  
f_m &=& \sum_{l}{f(l,m) },  \: m \in \L \cup \Lout . \label{4}  
\end{eqnarray}
Eqs \eqref{3}-\eqref{4} imply flow conservation at node $n$: number of vehicles entering and leaving $n$ is the same:
\begin{equation}
\sum_{l \in I(n)} f_l  = \sum_{m \in O(n)} f_m, \; n \in \N. \label{5}
\end{equation}
\textbf{Example 1} \ Figure \ref{fig8} shows a standard intersection with four approach and four departure directions.  So there are in all $4\times 3 =12$ turn movements, if U-turns are forbidden.  If there are $N$ intersections in the network, there are $12N$ flow or turn movements $\{f(l,m)\}$ in all, but only approximately $4N$ link flows $\{f_l\}$.  So fewer measurements are needed to estimate all link flows than to estimate all turn movements.

\noindent \textbf{Capacity constraint} \\
The number of vehicles $f(l,m)$ making movement $(l,m)$ is bounded by its capacity $s(l,m)$,  which is the product of its saturation flow rate and the green time or actuation duration,
\begin{gather}
f(l,m) \le s(l,m)  = \frac{1}{T_n}\sum_{i \in S_n} g_{n,i}(l,m) \times c(l,m), \quad \text{for all} \: (l,m). \label{6}
\end{gather}
In \eqref{6}, $T_n$ is the cycle time of the fixed-time controller at  intersection $n$, $S_n$ is the set of stages at $n$, $g_{n,i}(l,m)$ is the duration for which phase $(l,m)$ is actuated in stage $i$, and $c(l,m)$ is the saturation flow rate for phase $(l,m)$ at intersection $n$.
Phase  $(l,m)$ may be  actuated in more than one stage $i \in S_n$, which accounts for the sum on the right.

\textbf{Stability}\\
PointQ  is a discrete event simulation model of a continuous-time, delay-differential equation system that represents the traffic flow as a queuing network.
It is shown in \cite{FTControl} that \eqref{6} is necessary for the queueing network to be  stable and if the inequality in \eqref{6} is strict, the condition is sufficient for stability.

\section{Calibration} \label{sec-calibration}
\vspace*{-0.1in}
Suppose we have flow measurements (counts over say one hour) $\hat{f}_l$ on links $l \in M$, demands $\hat{d}_l$ at entry links $l \in D$, and turn ratios
$\hat{r}(l,m)$ for movements $(l,m) \in T$.  (The  $\: \hat{}  \:$  accent as in $\hat{f}_l$ denotes measured values.)  To minimize  the difference between the simulation values of the calibrated model and their measured values, we formulate a measurement equation:
\begin{eqnarray}
f_l &=& \hat{f}_l + {\epsilon_f}_l, \: l \in M ,\label{7}\\
d_l &=& \hat{d}_l  + {\epsilon_d}_l ,\; l \in D, \label{8}\\
f(l,m) &=& \hat{r}(l,m)f_l + {\epsilon_{f(l,m)}} , \; (l,m) \in T, \label{9}
\end{eqnarray}   
in which the $\epsilon_k$'s are measurement `errors'.
The calibration problem is 
\begin{gather} \label{10}
\min_{f_l,d_l,f(l,m)} \quad \sum_{l\in M}{\alpha_l {\epsilon_f}_l }^2+
\sum_{l\in D}{\beta_l  {\epsilon_d}_l }^2 +\sum_{(l,m)\in T}\gamma_{l,m} {\epsilon_{f(l,m)}} ^2 \\
s.t. \quad 0 \le f(l,m) \le s(l,m),  \mbox{ for all $(l,m)$}, \label{11}\\ 
f_l =  \sum_{m}{f(l,m) },  \: l \in  \L \cup \Lin  ,\label{12} \\  
f_m = \sum_{l}{f(l,m) },  \: m \in \L \cup \Lout . \label{13}  \\
\text{and equality constraints } \eqref{7}, \eqref{8}, \eqref{9}.\label{14}
\end{gather}
The decision variables are ${f_l,d_l,f(l,m)}$.  The weights $\alpha_l, \beta_l$ and $\gamma_{l,m}$  on the   errors  reflect confidence in the accuracy of the measurements. 
This is a quadratic programming problem.  All link and turn movement flows are taken as decision variables.  The problem is  feasible since the constraints are met with all decision variables equal to zero.

Let $f_l^*, d_l^*, f^*(l,m)$ be an optimum solution. Define the calibrated split ratios,
\begin{eqnarray} \label{15}
r^*(l,m)=\frac{f^*(l,m)}{f_l^*}.
\end{eqnarray}
\textbf{Remark} \
Suppose the measurements are error-free: all  $\epsilon_k = 0$.  As noted in Example 1, if there are $N$ standard nodes there are $4N$ link flows and
$12N$ turn movements. If we have too few measurements, the solution to the quadratic programming  problem will not be unique, which means that the available measurements (even if error-free) are insufficient to uniquely identify all link flows and turn movements. This raises several questions that are addressed in \S \ref{sec-indetermin}:
Which flows are uniquely identified and which are not? Which additional measurements will make a particular flow or all flows identifiable? Can we bound their accuracy even if some flows cannot be identified?

\section{Case study} \label{sec-case}
\begin{figure}[h!]
\centering
\includegraphics[width=6in]{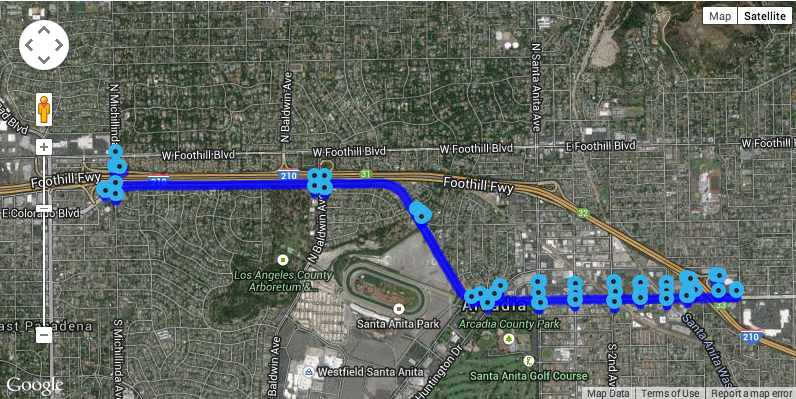}

\vspace*{0.5in}

\includegraphics[width=6in]{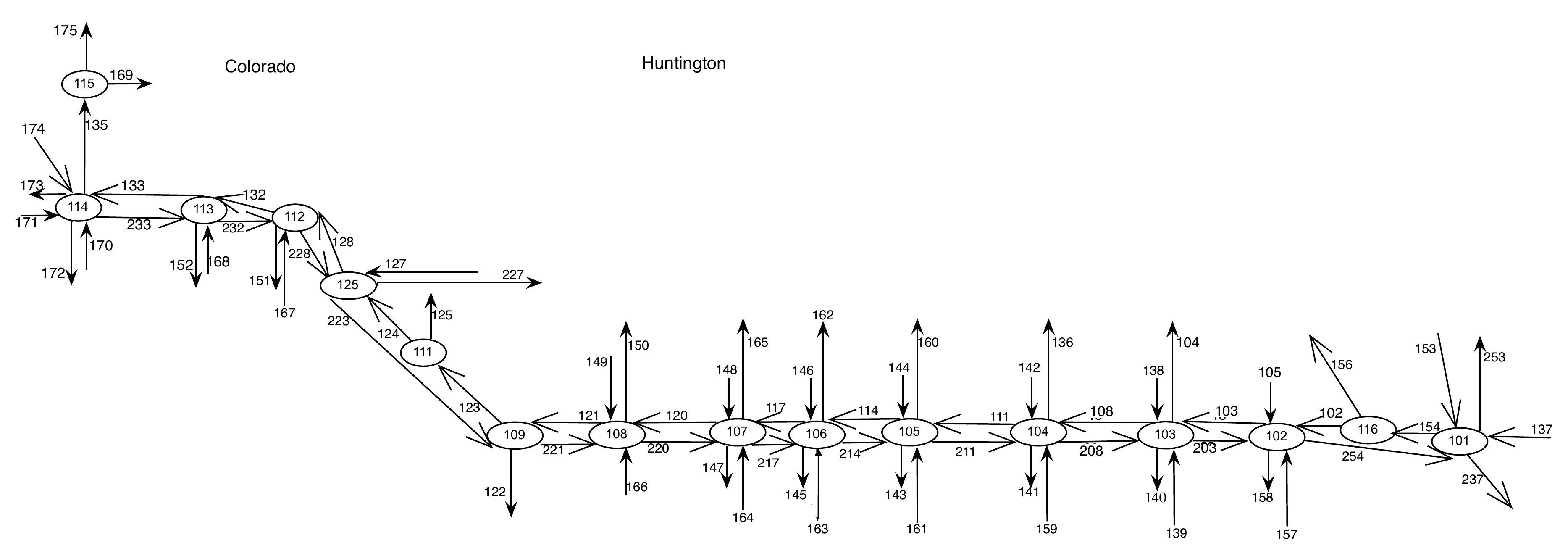}
\caption{Map of site and corresponding network graph.  Note: ID 105 means 10005.}
\label{fig-site}
\end{figure}

\vspace*{-0.1in}
We apply the calibration procedure to a section of the Huntington-Colorado arterial near I-210 in Los Angeles.  Figure \ref{fig-site} shows
a map of the site and its abstraction as a directed graph with 16 signalized intersections or nodes, 73 links, and 106 turn movements.  This gives 179 flows in all.  Flows in 30 links and 57 turn ratios are measured, many flows are unmeasured as is common in practice.

The table in the Appendix  gives the calibration results obtained by solving \eqref{10}-\eqref{14}.  The table lists link IDs, measured flows $\{\hat{f}_l\}$, and calculated flows $\{f_l\}$ and the movements (from $l$ to $m$) on the right hand side and the measured turn ratios $\{\hat{r}(l,m)\}$ and calculated turn ratios $\{r(\ell, m)\}$ on the left. Flows are in vehicles per hour (vph).  For example, in link 200054 (abbreviated as link 254 in Figure \ref{fig-site}) the measured flow is 1,016 vph while the optimization calculation yields 1,014 vph. Also, for movement
(100037, 100054) both measured and calculated turn ratios equal 0.65. In the table an entry of `-1' means that the corresponding measurement is not available.
The table shows that the calibrated model  calculations  are close to measured values.  The small errors are the result of  (i) the relatively large number of unmeasured quantities, which make it easier to `fit' the model \eqref{11}-\eqref{14}, and  (ii) consistency of the available measurements.

The quadratic program uses data for a single  hour.  Since the flows are considered to be constant during this period
and the travel time across the network is much smaller than one hour, it is appropriate to assume the steady state model.  If we have data
for several time periods the exercise can be repeated and the mismatch between measured and simulated value should be small again.
One may believe that the calibrated turn ratios $r^*(l,m)$ should be  the same if (say) measurements are made for the same hour over several days.  We can attempt to enforce this by taking $r^*(l,m)$ obtained from calibration for one day and use it as a `measured' turn ratio $\hat{r}(l,m)$
for other days.  A more direct way would be to make $r(l,m)$  a \textit{variable} and add the constraint:
\[f(l,m) = r(l,m) f_l,\]
but this is a \textit{nonlinear} constraint, and computationally more difficult to handle.

We will refer to the calibrated demands, turn ratios and link flows in the Appendix as  values of the \textit{baseline} model.

\subsection{Simulation} 
A PointQ model  is specified by the five items listed at the beginning of \S \ref{sec-model}.  The network graph is obtained from a map like the one in Figure \ref{fig-site}.  Saturation flow rates (Item 2) are obtained from high-resolution measurements as in \cite{ITSC2014} or as default values  from the Highway Capacity Manual (\cite{HCM}), as in the case study.  Item 5 is obtained from the  traffic departments of  cities sharing the arterial network.  Some link flows, demands and turn ratios are obtained from special data collection efforts  and  from  controller-based counts.  Formulating and solving the calibration problem \eqref{10}-\eqref{14} yields the fully specified  baseline model.  We can now run that model and obtain  performance measures for comparison with other field measurements (like route travel times) to validate or reject the model.  We give four examples.

\begin{figure}[h!]
\centering
\includegraphics[width=6.2in]{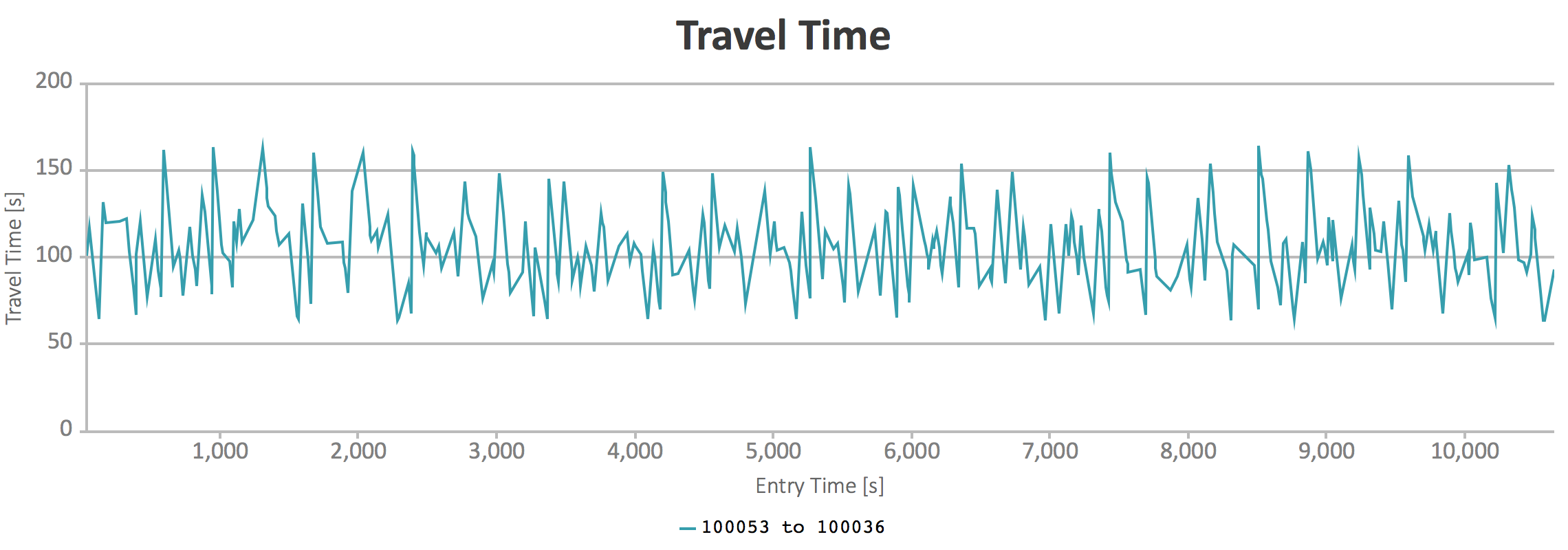}
\caption{Travel time of 363 vehicles entering link 100053 and departing from link 100036.}
\label{fig:TT-CM}
\end{figure}

Figure \ref{fig:TT-CM} displays the travel time of the 363 individual vehicles that enter the network at link 100053 and leave it at link 100036
(see Figure \ref{fig-site} for the location of these links).  The mean and standard deviation of these 363 travel times are
\begin{equation}
\mu = 107\mbox{s and $\sigma$ = 24s}.\label{16}
\end{equation} 
If for example we had travel time measurements using (say) Bluetooth readers, we could use those measurements to test the null hypothesis that parameters of the travel time distribution agree with \eqref{16}.   Note that the
entire travel time plot lies within $[\mu-2\sigma, \mu + 2 \sigma]$, so a statistical test of that hypothesis would be  that the travel time field measurements lie in this range as well.

\begin{figure}[h!]
\centering
\includegraphics[width=6.2in]{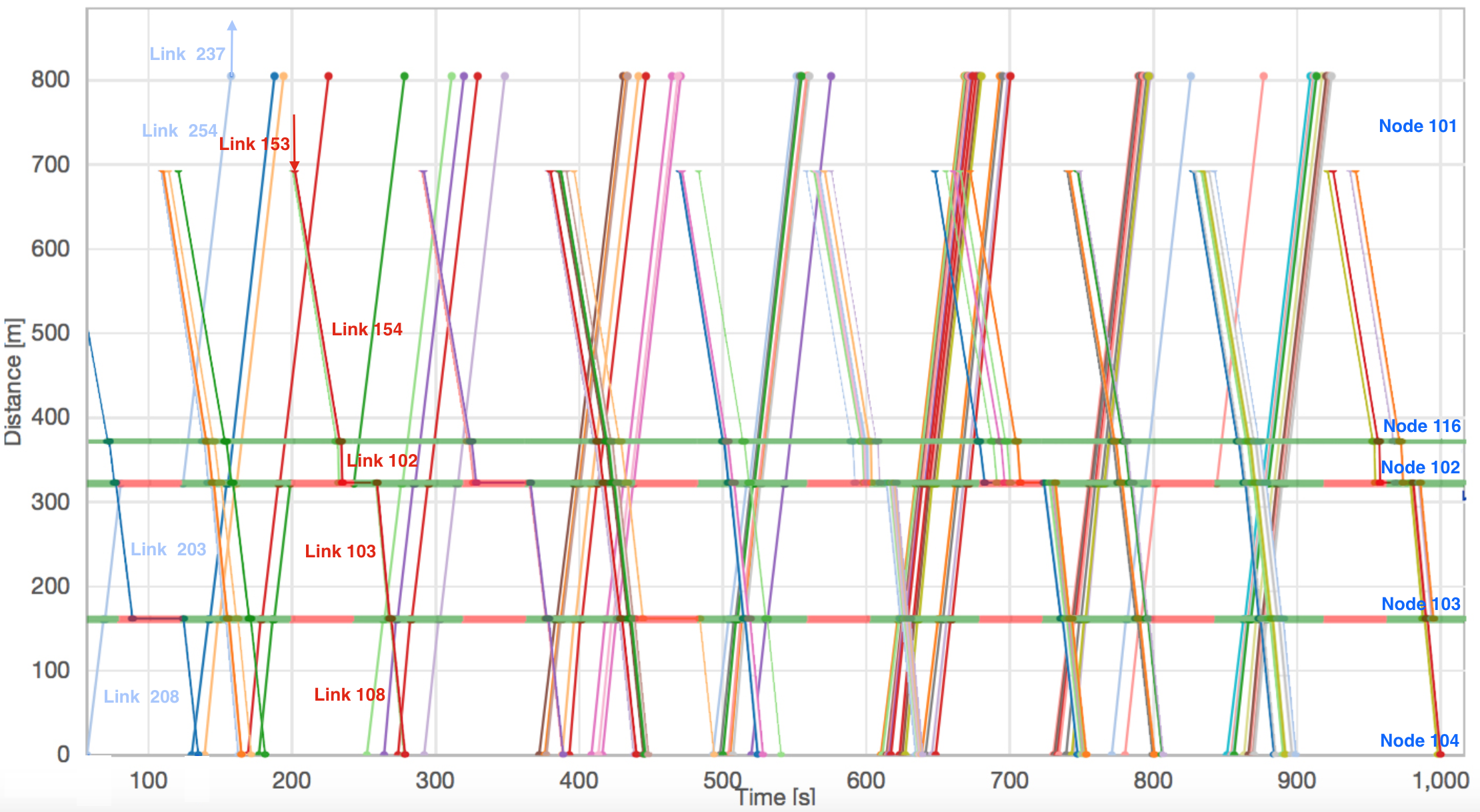}
\caption{Trajectories of vehicles that traverse links 100053-10008 and 20008-200037.}
\label{fig:traj}
\end{figure}

Figure \ref{fig:traj} gives the trajectories of all 43 and 47 vehicles that traverse  links 100054 and 10008 in the westbound direction and links 20008 and 200037 in the eastbound direction.   Also shown are the red/green durations for the phases at the intersections   encountered by these trajectories.\footnote{There are trajectories that traverse link 103 and stop at node 102  even during green, which seem incorrect, but it is because of other vehicles in front that make a turn and do not continue
on link 154, so their trajectories are not displayed in Figure \ref{fig:traj}.}
Too many vehicles appear to stop at the end of the short link from node 100016 to node 10002 suggesting a need to redesign the offset. Direct field observations may validate or reject  the signal progression quality for this route as indicated by the figure.
\begin{figure}[h!]
\centering
\includegraphics[width=6.2in]{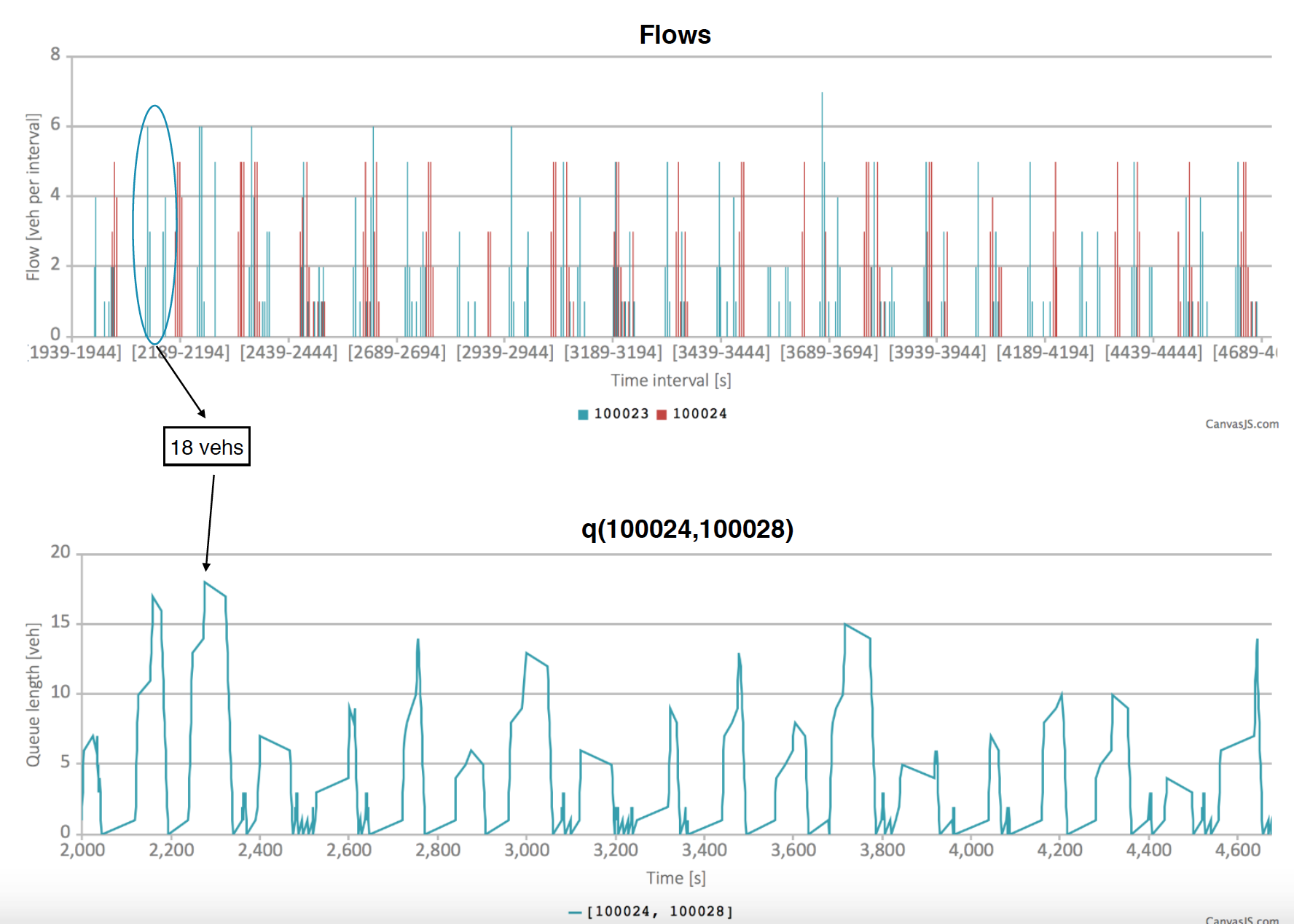}
\caption{Queue $q(100024,100028)$ of the through movement at intersection ID 100025.}
\label{fig:queue}
\end{figure}

Figure \ref{fig:queue} uses the PointQ output for a more detailed analysis of $q(100024,100028)$.
The queue  grows as vehicles arrive in link 100024 and shrinks as they discharge into link 100028.  
The flow on a link within (say) a 5s interval is calculated as the number of vehicles that \textit{leave} this link during that interval.  The upper part of
Figure \ref{fig:queue} shows 5s flows leaving links 100023 and 100024. 
From Figure \ref{fig-site} we see that vehicles leaving 100023 enter 100024 and those leaving 100024 enter 100028.   Thus surges in the
flow on 100023 lead to surges in $q(100024,100028)$ and surges in flow on 100024  lead to drops in this queue.  This is illustrated in the figure by
the departure of a 18 vehicle-platoon from link 100023 followed by a 18-vehicle increase in the   size of $q(100024, 100028)$.  Another observation that corroborates the internal consistency of the simulation model is that  the vehicle platoons leaving link 100024 are separated by 145s, which is the cycle time of intersection 100025.  Lastly, on-site observation may confirm the simulation prediction that the queues in this intersection clear in every cycle. 

\begin{figure}[h!]
\centering
\includegraphics[width=6.2in]{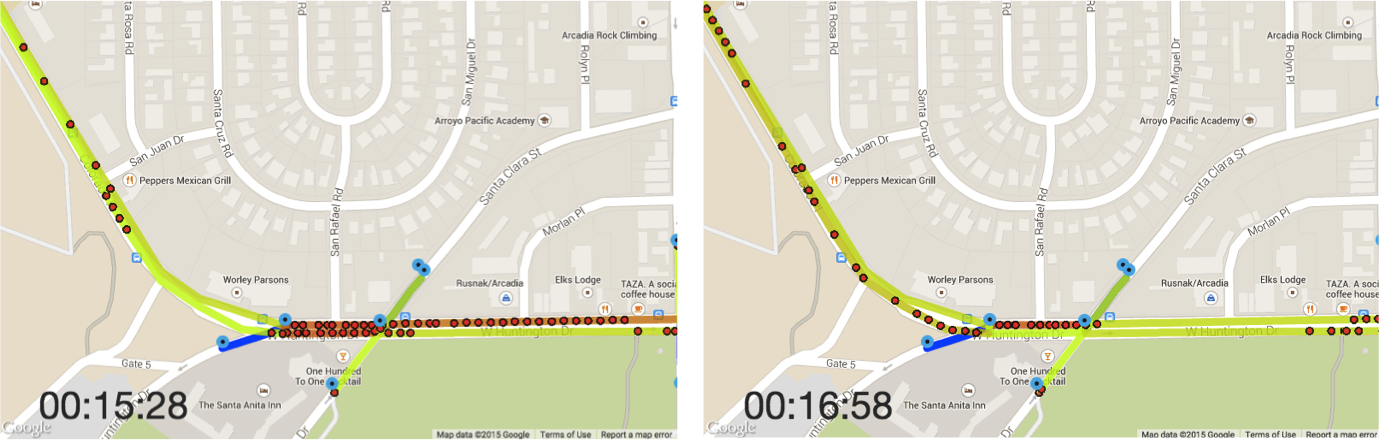}
\caption{Two frames of visualization of vehicle movement.}
\label{fig:animate}
\end{figure}
As noted earlier, PointQ has a facility to visualize the movement  of vehicles like other microsimulation packages.  Figure
\ref{fig:animate} shows two frames of an animation of the movement of all vehicles, each represented by a small circle.  This helps one understand the formation of queues and the passage of vehicles through an intersection in platoons.  It also aids in appreciating how congestion propagates under heavy demand, see \S \ref{sec-macro}.

\section{Experiments} \label{sec-exp}
\vspace*{-0.1in}
Assuming the calibrated baseline model is satisfactory, we use it to conduct four numerical experiments.
In Figure \ref{fig-site}, entry link 100053 is fed by the off-ramp from I-210W and link 100069 feeds into the on-ramp, downstream of the off-ramp.
The first two experiments estimate how much additional flow can be diverted from I-210W onto the arterial and then returned to the freeway.  The objective is to
evaluate the potential of diversion onto the arterial in the event of a capacity-reducing incident on the highway.  The third experiment compares travel times along the diversion route for two fixed-time controls and with max pressure control (\cite{MPtrc}).  The fourth experiment compares a single-commodity and
a multiple-commodity model that both give the same link flows.

\subsection{Simple diversion}  \label{sec-simple}
\vspace*{-0.1in}
We  take as baseline traffic the calibrated demands, turn ratios, and flows.   We augment this traffic with a
 `diversion'  of $D$ vph from the freeway,   starting in link 100053,  ending in link 100069, and following the direct route through links 
\[100054, 100012, \cdots, 100032, 100035 \mbox{  (see Figure \ref{fig-site})}.\] 
We want to find the largest diversion $D$ that can be accommodated in \textit{addition} to the baseline traffic with the \textit{same} timing plans.
Let $T_D$ be the set of turn movements 
\[(100054, 100012), (100012, 100013), \cdots, (100033,100035), (100035,100069)\] 
followed by the diverted traffic.
Let $\{f^*(l,m)\}$ be the turn movements of the baseline traffic obtained as the solution of \eqref{10}-\eqref{14}.  Then the largest diversion is the extra  flow that can be accommodated, namely
\begin{gather} \label{17}
D^* = \max_D  D \\
\mbox{s.t. } f^*(l,m) + D \le s(l,m) , \: (l,m) \in T_D. \label{18}
\end{gather}
Clearly,
\[D^* = \min \{s(l,m) - f^*(l,m)~|~ (l,m) \in T_D\}.\]
We find $D^*= 270$ vph.    This can
be compared with the baseline flow of 522 vph on the off-ramp  (link 100053).   The mean and standard deviation of the travel times of vehicles in the baseline traffic that traverse links 100053 and
100036 now is
\begin{equation}
\mu = 107 \mbox{s  and  $\sigma = 23$s}, \label{19}
\end{equation}
which, upon comparing with \eqref{16},  shows that this diversion does \textit{not} deteriorate the travel  of the baseline traffic.
However, with peak traffic on I-210W of 10,000 vph, the diversion of 270 vph does not by itself offer the prospect of significant relief, although it can be an element of a response to a delay-causing incident that includes ramp metering and variable message sign alerts.

\subsection{Diversion with re-timing} \label{sec-retime}
\vspace*{-0.1in}
We now find the largest diversion $D^+$ that can be accommodated in addition to the baseline traffic when we are allowed to \textit{change}
the timing plans.  This means that in \eqref{6} the $\{g_{n,i}(l,m)\}$ are now decision variables instead of being specified
constants.  In place of \eqref{17}-\eqref{18}, we have the following linear programming problem:
\begin{gather} \label{20}
D^{+*}=\max_{D, g_{n,i}(l,m)} D \\
\mbox{s.t. } f^*(l,m) + D \le s(l,m) , \: (l,m) \in T_D, \;\;  f^*(l,m) \le s(l,m) ,  \: (l,m) \not\in T_D, \label{21} \\
s(l,m) =  \frac{1}{T_n}\sum_{i \in S_n} g_{n,i}(l,m) \times c(l,m) , \mbox{ for all } (l,m) ,\label{22}\\
\sum_{i \in S_n} \sum_{(l,m)} g_{n,i}(l,m) \le T_n - L_n, \mbox{ for all } n ,\label{23} \\
g_{n,i}(l,m) \ge 0, \mbox{ for all } (l,m), n .\label{24}
\end{gather}
The difference between \eqref{17}-\eqref{18} and \eqref{20}-\eqref{24} is that in the latter  the green duration for each movement is a decision variable, subject to the constraint that the total green time at intersection $n$ is bounded by the cycle time $T_n$ minus the lost time $L_n$.
The maximum value in \eqref{20} turns out to be $D^{+*} = 756$ vph, almost three times the maximium diversion without retiming of 270 vph.  It may seem surprising that  the mean and standard deviation of the travel times of vehicles in the baseline traffic that traverse links 100053 and
100036 now is \textit{decreased} (compare \eqref{19}):
\begin{equation}
\mu = 83.3 \mbox{s  and  $\sigma = 16.7$s}. \label{25}
\end{equation}
The improved performance is  due to the fact that in this experiment the durations are selected optimally
according to \eqref{20}-\eqref{24}, so the V/C ratio along this route is decreased with the same cycle time.  Clearly diversion with optimal re-timing offers more relief to the freeway congestion.

\subsection{Max pressure control} \label{sec-MP}
\vspace*{-0.1in}
Max pressure (MP)  is a distributed signal control scheme that adjusts the green durations at an intersection as a function only of the lengths of the queues immediately upstream and downstream of the intersection.  An MP control is parametrized by the number of decision times within each cycle that the duration can be adjusted.    The theory of MP is developed in \cite{MPtrc}, where it is shown that the queue length resulting from MP control is inversely proportional to the number of decisions per cycle.  Simulation studies of MP are presented in \cite{wodes} and \cite{liorisTRR}.
 \begin{figure} [h!]
\centering
\includegraphics[width=3.2in]{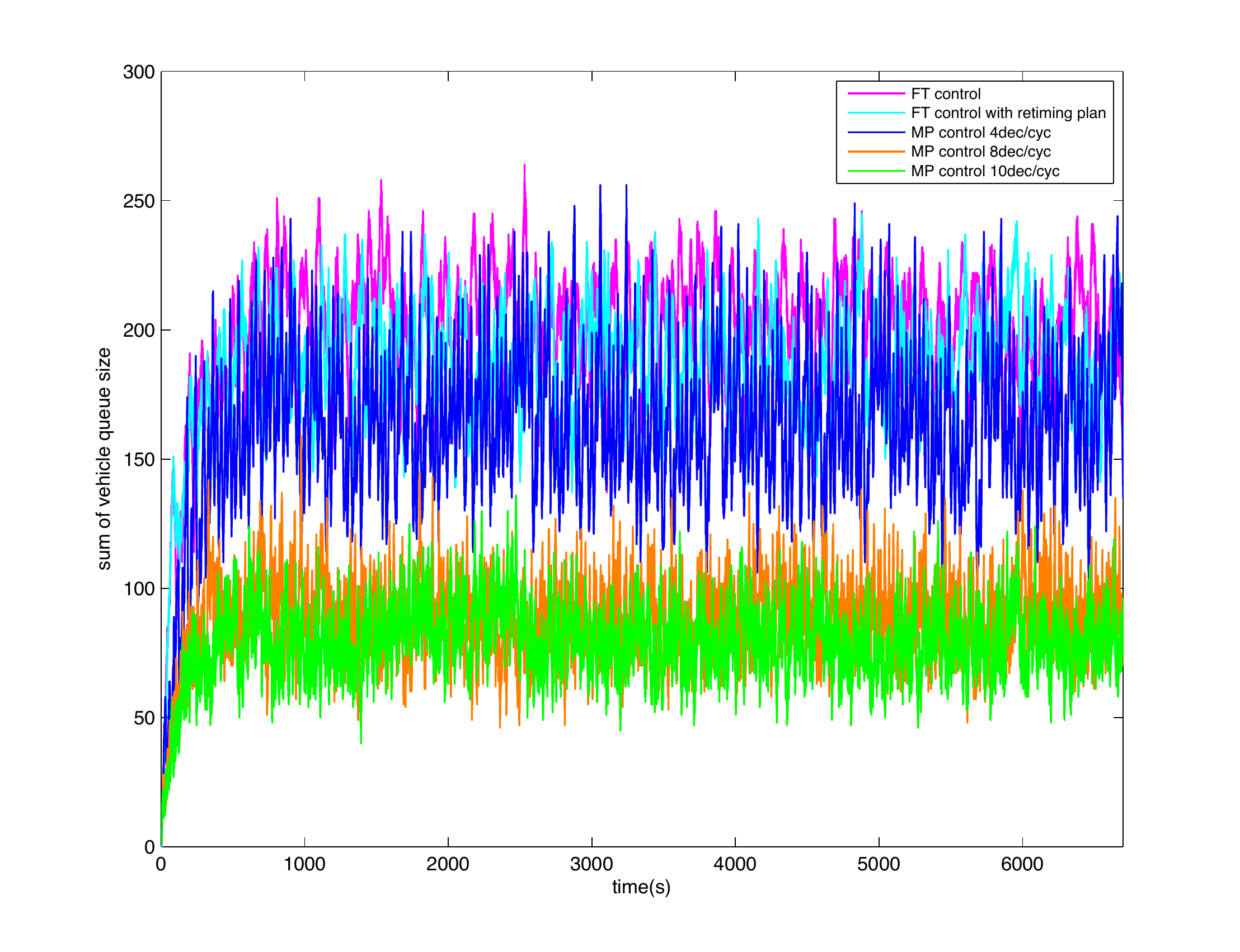} \includegraphics[width=3.2in]
{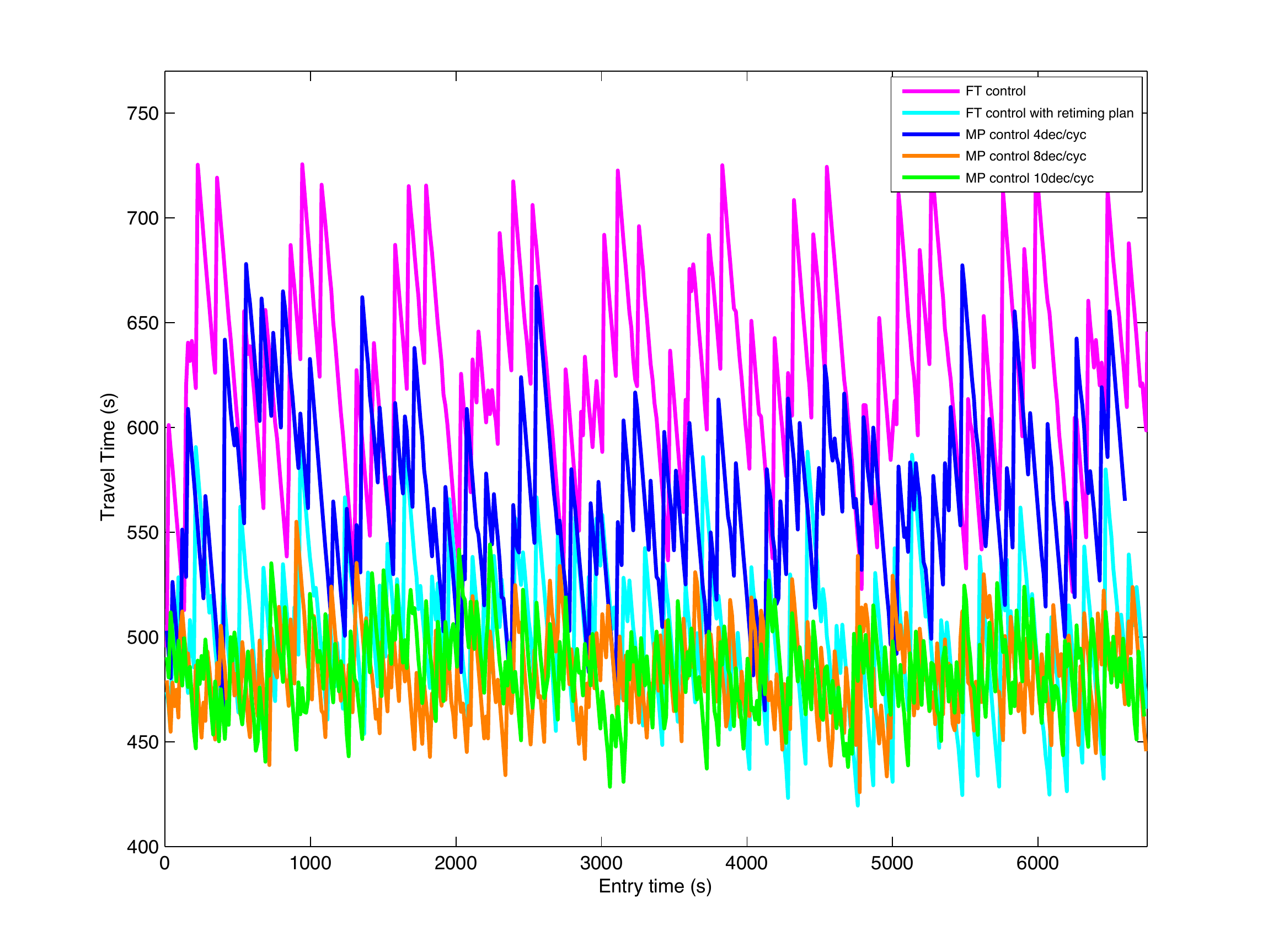}
\caption{Sum of all queues (left) and travel time from off-ramp to on-ramp (right) for the baseline traffic, FT vs MP.}
\label{fig:TT}
\end{figure}

Figure \ref{fig:TT} shows the sum of all queue lengths (left) and the travel time (right) from off-ramp (link 100053) to on-ramp (link 100069) for the baseline traffic with an additional flow of 270 vph diverted from the freeway, for five different signal controls: 
the fixed time (FT) baseline control of \S \ref{sec-simple}, FT control with re-timing of \S \ref{sec-retime}, and MP control with 4, 8, 10 duration decisions per cycle. The plots clearly demonstrate the benefits of MP relative to FT control: MP with 10 decisions vs  FT has an average total queue size of 75 vs 200 vehicles, and off-ramp to on-ramp travel time of 475s vs 650s on average, which is a travel time reduction of 27 percent.  Since queuing delay is roughly proportional to the square of the queue size, MP with 10 decisions has an intersection delay of $(75/200)^2$ or 14 percent of FT delay, that is, it achieves a queuing delay reduction of 86 percent compared with the baseline.  Note that the travel time along the diversion route for the FT control with re-timing is similar to,
but the sum of queues is much larger than,  that for MP with 10 decisions, which may be expected since the re-timing is designed to facilitate movement along the diversion route.

\begin{figure} [h!]
\centering
\includegraphics[width=6in]{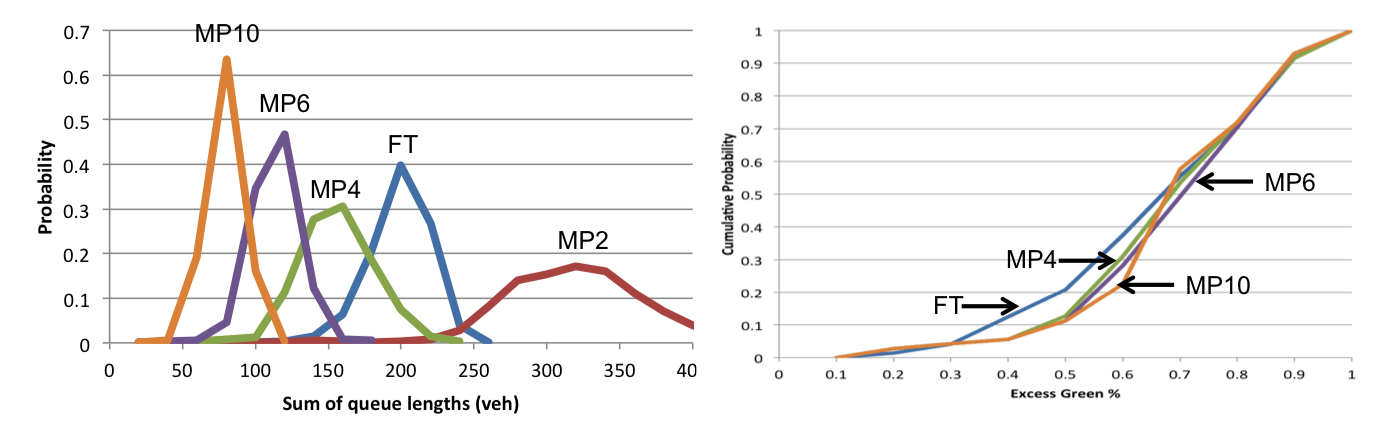}
\caption{Histogram of sum of all queues (left) and cumulative probability distribution of `excess green' (right) under FT and various MP controls.}
\label{fig:cumsum}
\end{figure}
The plots of Figure \ref{fig:cumsum} give a different perspective on the relative performance of FT vs MP control.  The plot on the left is the histogram (drawn as a continuous curve rather than a bar chart) of the sum of all queues under various controls.  Thus it is simply the stationary distribution of the time series in the left of Figure \ref{fig:TT}, and shows more clearly the advantage
of MP over FT and the advantage of more decisions per cycle.  The plot on the right uses the notion of `excess green' $e (\phi)$ of a phase $\phi$  defined as the fraction of time that $\phi$ is actuated but serves no vehicle, i.e.
\[e(\phi) =  \frac{\mbox{time that $\phi$ is actuated and its queue is 0}}{\mbox{time that $\phi$ is actuated}}.\]
The cumulative distribution function $e \mapsto F(e)$ is plotted on the right:
\[F(e) = \frac{\mbox{number of phases $\phi$ with $e(\phi) \le e$}}{\mbox{total number of phases $\phi$}}.\]
$F(e)$ is a measure of excess capacity for a given intersection control.  Thus under FT (MP) 80 (90) percent of the phases have an excess green of at least 50 percent, because MP clears queues more quickly than FT.
\subsection{Single vs.\ multiple commodity}
\vspace*{-0.1in}
In the  experiments of \S \ref{sec-simple} and \S \ref{sec-retime}  vehicles  belonging to the baseline flow and those being diverted follow routes specified by different turn ratios.  So this is a 2-commodity flow.  More generally, a $P$-commodity flow has $P$ different demands, link flows, turn movements and turn ratios, $\{d^p_l, f_l^p, f^p(l,m), r^p(l,m)\}$,
$p=1, \cdots, P$, each satisfying its own version of the constraints \eqref{6}-\eqref{10}.  The aggregate turn movements and link flows are
$\{f(l,m) = \sum_p f^p(l,m)\}$, $\{ f_l = \sum_p f^p_l \}$.  These aggregate quantities define a single commodity flow with (non-unique) turn ratios:
\begin{equation}
r(l,m) = \frac{\sum_p [r^p(l,m) f^p_l]}{\sum_p f^p_l}.\label{26}
\end{equation}
Thus as far as link flows are concerned, one can always use a single-commodity flow model.  This transformation into a single commodity model allows one to use a simple simulation model like CTM  as in \cite{lo2001}.   But note that using \eqref{26} requires knowing each of the $P$ commodity flows.  Moreover the aggregate turn ratios $\{r(\l,m)\}$ will change if any of the component flows changes, so simulating  time-varying  multi-commodity demands with a single commodity demand is virtually impossible.  Finally, the \textit{routes} followed by vehicles in the aggregate single-commodity model bear little relation to the routes in the $P$-commodity model.

\section{Identifiability} \label{sec-indetermin}
\vspace*{-0.1in}
As remarked at the end of \S \ref{sec-calibration}  often in practice not enough flows are measured  for one to identify or impute all link and turn movement flows.  We now give   conditions on the network graph that determine whether or not a  particular flow can be calculated from  the available measurements.  We first consider link flows and then extend  the results to turn movements.  It is assumed that the measurements are error-free.  

\begin{figure}[h!]
\centering
\includegraphics[width=5in]{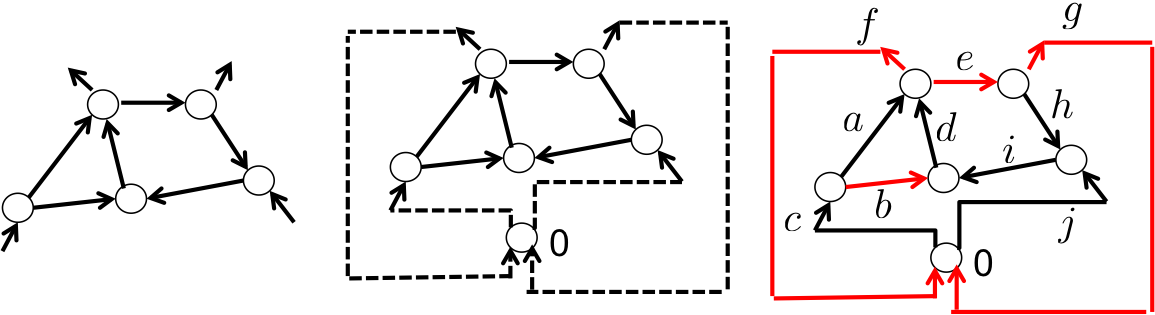}
\caption{Original graph (left), adding node 0 (middle), red links are measured (right).}
\label{fig:identify}
\end{figure}

Let $\G = (\N, \E)$ be a directed graph with nodes $\N$ and directed links $\E$.  It is assumed that $\G$ is strongly connected, that is for any two nodes $n, m$ there is a directed path from $n$ to $m$ and another directed path from $m$ to $n$.  The graph of Figure \ref{fig-site} is \textit{not}
directly connected, since exit links are not connected to entry links.  So we add an artificial node, node 0, and have every entry link start at 0 and every exit link end at 0.  Figure \ref{fig:identify} illustrates the procedure.  The original network graph is shown on the left.  It has 5 nodes and 10 links, including 2 exit and 2 entry links. Adding node 0 (at the bottom) and connecting (by dashed lines) the entry and exit links to 0 gives the strongly connected graph in the middle.  Suppose the flows in the four red links $(b,e,f,g)$ in the graph on the right are measured and the flows in the remaining six black links are not measured.  By Theorem \ref{T1} below the red measurements together with the flow conservation equation \eqref{5} uniquely determine all black link flows.

We will say that a sequence of links $(\l_1, \cdots, \l_n, \l_{n+1}=\l_1)$ forms an (undirected) \textit{cycle} if they form a cycle in the ordinary sense when we ignore their      direction.  The sequence forms a \textit{directed} cycle if all links have the same  direction.  
In Figure \ref{fig:identify} $( b, d, a, b)$ and $(j, h, g, j)$ are  cycles,  whereas $(a, f, c, a)$ and $(f, c, b, d, f)$ are directed cycles.

\begin{theorem}\label{T1}
 Let $\G = (\N, \E)$ be a strongly connected directed graph.  Suppose  flows on the subset  $\M$ of links are measured
and flows on the remaining links  $\U = \E \setminus \M$ are unmeasured.  The flow on  link $e \in \U$ can be determined from the measurements on $\M$ and the
conservation flow equations \eqref{5} if and only if $e$ does not belong to a  cycle comprised of links from $\U$. 
\end{theorem}

\textbf{Proof}  As in Figure \ref{fig:identify} paint the measured links red and the unmeasured links black.  Suppose $e$ is a black link that does not belong to any cycle of black links.   

We first prove sufficiency.
Suppose link $e$ goes from node $i$ to $j$.  Let $B$ be the set of all nodes $b$ except $j$ for which there is an undirected path 
from $i$ to $b$ consisting only of black links.  Let $R = \N \setminus B$ be the remaining nodes.   Every link between $B$ and $R$ except $e$ must be red, that is it is measured. Hence the set $C$ of all links
that leave or enter $B$ together with link $e$ forms a cutset that separates $B$ and $R$.  By flow conservation, the algebraic sum of flows through the cutset
must be 0, so
\[\sum_{\l \in C} f_l + f_e = 0,\]
and since the flows in $C$ are measured we obtain $f_e = -\sum_{\l \in C} f_l $.  In this equation we abuse notation and assign a sign (+ or -) to $f_l$ depending on whether link $l$ is pointing into or away from $B$.

To prove necessity let $L= (l_1, \cdots, l_n = \l_1 )$ be a cycle of black links.  It is enough to find two different sets of flows $ \{f_{\l}^1\}$ and $\{f_{l}^2\}$ such that  $f_l^1 = f_l^2, l \notin L$ and $f_l^1 \neq f_l^2, l \in L$.  Let $\{f_l^1\}$ be any flow vector with all components  strictly
larger than some $\epsilon > 0$.  Because the graph is strongly connected, one can always find such a flow vector.  Now select a positive direction
for links in the cycle $L$.  Construct $\{f_l^2\}$ by
\[ f_l^2 = \left \{
\begin{array}{ll}
f_l^1 + \epsilon, & \mbox{ if $l \in L$ and the direction of $\l$ is postive } \\
f_l^1 - \epsilon, & \mbox{ if $l \in L$ and the direction of $\l$ is negative } \\
f_l^1, & \mbox{ if $\l \notin L$}.
\end{array}
\right .
\]
It is easy to see that the conservation equation  \eqref{5} is satisfied. Furthermore, $f_l^2 > 0$ for all $l$, so this is a legitimate flow. \hfill $\Box$
\begin{corollary}\label{C1}
Suppose $\G$ has $l$ links and $n$ nodes.  Then $l-(n-1)$ is the minimum number of links whose flow must be measured in order to determine the flows on all links of $\G$.
\end{corollary}
\textbf{Proof}  Every tree of $\G$ has $n-1$ links, and each of the non-tree links is part of a unique cycle  involving only tree links.  So every non-tree link must be measured to prevent a cycle comprised only of unmeasured links.  The number of non-tree links is $l-(n-1)$. \hfill $\Box$

$\G$ may have many trees, each of which yields a minimal set of link measurements.

\begin{corollary}\label{C2} If there is a directed cycle of unmeasured links the unmeasured flows can be arbitrarily large.
But if there is no directed cycle of unmeasured links the unmeasured flows can be bounded from the measured flows.
\end{corollary}
 \textbf{Proof}  If the cycle $L$ in the proof of Theorem \ref{T1} is directed, i.e.\ all links are either in the positive (or negative) direction, then $\epsilon$ ($- \epsilon)$ can be arbitrarily large.  To prove the second assertion, note that the subgraph $(\H, \U)$ of all unmeasured links  is a directed acyclic graph, as it has no directed cycles.  Let $\prec$ be a partial order on $\H$ with $h \prec h'$ if there is a directed path from $h$ to $h'$.  Let $h_1 \prec h_2 \prec \cdots \prec h_K$
 be a total order on $\H$ compatible with the partial order.  Partition the links in $\G$ that are incident to $h_1$ as $\M_1 \cup \U_1$ with
 $\M_1 \subset \M$ and $\U_1 \subset \U$.  By \eqref{5} at node $h_1$
 \[\sum_{l \in \M_1} f_l + \sum_{l \in \U_1} f_l = 0  \mbox{ (algebraic sum)}, \]
 which gives the bound
 \[|f_l| \le |\sum_{l \in \M_1} f_l | , l \in \U_1.\]
 We proceed by induction.  Suppose we have bounds on flows on links incident to $h_1, \cdots, h_{k-1}$.   Partition the links incident to $h_k$ as
 $\M_k \cup \U_k$ with $\M_k$ consisting of the links that are in $\M$ or incident to $h_1, \cdots, h_{k-1}$.  By \eqref{5} at node $h_k$
 \[\sum_{l \in \M_k}f_l + \sum_{l \in \U_k}f_l = 0 \mbox{ (algebraic sum)}, \]
 which gives the bound
 \[|f_l | \le |\sum_{l \in \M_k} f_l | , l \in \U_k. \hfill \Box\]

\subsection{Measuring turn movement flows}
\begin{figure}[h!]
\centering
\includegraphics[width=4in]{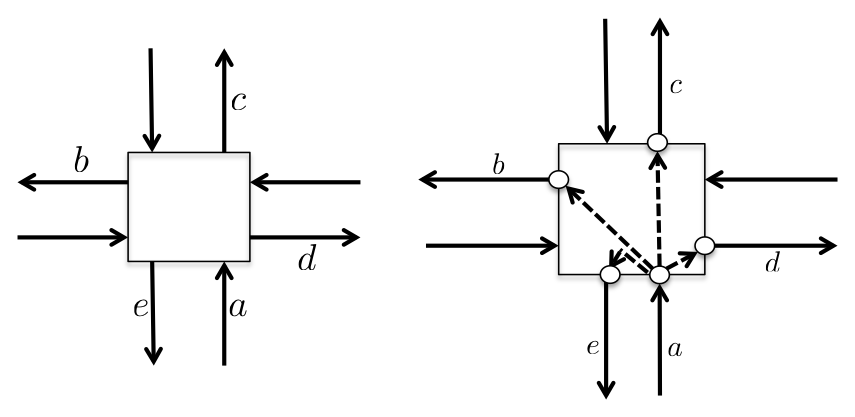}
\caption{Original node (left), adding turn movements (right).}
\label{fig:turnmovement}
\end{figure}

In the preceding section we considered  the link flows  only and ignored turn movement flows $f(l,m)$  that may be measured or need to be imputed.  But this deficiency is easily corrected by adding a link $(l,m)$ for every turn movement we want to consider and take $f(l,m)$ as the flow on that link.  This is illustrated in Figure \ref{fig:turnmovement}.

On the left in Figure \ref{fig:turnmovement} is a node with four input links and four output links.  On the right we add four additional links corresponding to the movements $(a, b), (a, c), (a, d)$ and $(a,e)$. These links carry the flows $f(a,b), f(a, c),f(a,d), f(a,e)$.  So  to consider these turn movement flows we can augment the graph, and the results above will hold for this augmented graph.  Note that the movement $(a,e)$ is a U-turn.  If this is prohibited, we simply delete this link or regard it as a measured link with flow $f(a,e) = 0$.  

\subsection{Known turn ratios} 
\begin{figure}[h!]
\centering
\includegraphics[width=4in]{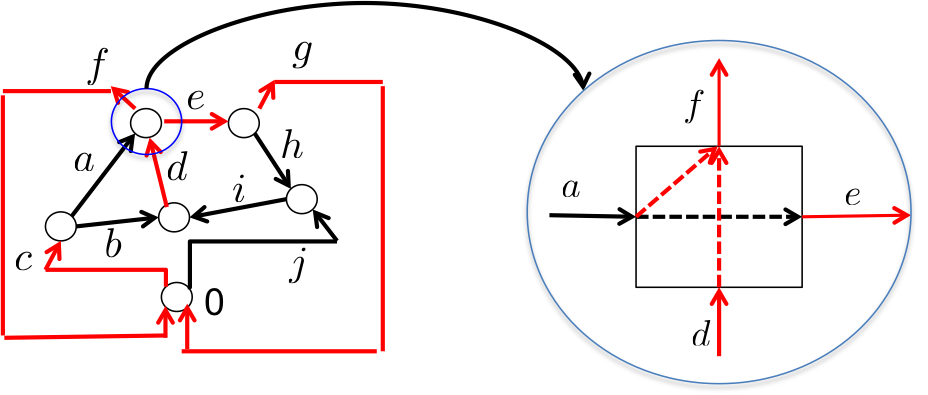}
\caption{Turn ratios for red (black) dashed links are known (unknown).}
\label{fig:turnratio}
\end{figure}

To consider a measured or unmeasured turn movement \textit{flow}  $f(l,m)$ we add the link $(l,m)$.  Suppose, however, that the turn \textit{ratio} $r(l,m)$ is measured but the flow $f(l,m)$ is not measured.  How can we use this information?  As an example, suppose the turn ratio $r(a,f)$ in Figure \ref{fig:turnratio} is known.  Suppose further that the link $(a,f)$ is not part of any unmeasured cycle.  Then the flow $f_a$ can be obtained and hence
$f(a,b) = r(a,b) f_a$ can be calculated.   In the example of Figure \ref{fig:turnratio} suppose the turn ratios $r(a,f)$ and $r(d,f)$ are known.  Then
we can obtain $f_a = [f_f - r(d,f)f_d]/r(a,f)]$ and $f(a,e) = f_a - r(a,f)f_a$.  Then there is no cycle with unmeasured links and so all flows can be calculated.
This argument leads to the following sufficient condition.
\begin{corollary} \label{C3}
Suppose link $l$ has known turn ratios $r(l,m_1), \cdots, r(l,m_k)$ $[\sum_i r(l,m_i) = 1, r(l,m_i) > 0]$.  Suppose link $(l,m_i)$ is not part of any cycle of unmeasured links.  Then all turn movement flows $f(l, m_1) , \cdots, f(l,m_k)$ can be determined.
\end{corollary}

\subsection{Estimating VMT} \label{sec-vmt}
If the unmeasured links (including turn movement links if needed) contain no directed cycle, Corollary \ref{C2} implies that flows in the cycles can be bounded.  A simple way of computing the inaccuracy 
in the traffic in the unmeasured cycles is to calculate upper and lower bounds on the vehicle miles traveled (VMT) on these cycles.  This is of independent interest as VMT has become an important performance measure.\footnote{California Senate Bill 743 states that the criteria for determining the significance of transportation projects may include VMT and but \textit{not} vehicle delay in assessing environmental quality.}  VMT bounds can also be used to assess the inaccuracy of the calibration procedure.  If all link flows $\{f_l\}$ are measured and the length $d_l$  of every link is known, 
\[VMT = \sum_l d_l f_l.\]

We want to estimate VMT when not every link flow is measured.
As in Theorem \ref{T1} suppose $\G=(\N, \E)$ is strongly connected,  flows on $\M \subset \E$ are measured, and flows in $\U = \E \setminus \M$
are unmeasured.  Then
\begin{equation} \label{27}
VMT = \sum_{l \in \M} d_l f_l + \sum_{l \in \U} d_l f_l .
\end{equation}
The first term on the right can be calculated from measurements, and we estimate the second term.

Express the flow conservation relations \eqref{5} in matrix form as 
\begin{equation}
Af = 0 ,\label{28}
\end{equation}
in which $f$ is the flow column vector with entries $f_l, l \in \E$, $A$ is the node-link incidence matrix with entries $A(n,l), n \in \N, \l \in \E$:
\[A(n,l) = \left \{
\begin{array}{ll}
+1, \mbox{ if } l \in O(n) \\
-1, \mbox{ if } l \in I(n) \\
0, \mbox{ else}. \; 
\end{array}
\right .
\]
Partition $f = (f^m, f^u)$ with entries in $\M$ and $\U$ and  $A = [A^m ~|~ A^u]$ so \eqref{28} can be written
\begin{equation} \label{29}
A^uf^u = b = -A^m f^m.
\end{equation}
Then $b$ is known since $f^m$ is measured.  Suppose $f^u$ is of dimension $k$, the number of unmeasured links.  Then $f^u$ is uniquely determined by \eqref{29} if and only if
rank$(A^u) = k$, and by Theorem \ref{T1}  rank$(A^u) < k$ if and only if the set of links $\U$ contains a cycle.

Suppose $\U$ contains one or more cycles.  Then the second term in \eqref{27} is not fully determined from the measurements, but an upper and lower bound on this term is given as the solution to two linear programs:
\begin{gather} 
VMT^u_+ =\max_{f^u}  \sum_{l \in \U} d_l f_l^u  \label{30}\\
\mbox{s.t. } A^uf^u = b, f^u \ge 0, \label{31}
\end{gather}
and 
\begin{gather} 
VMT^u_- =\min_{f^u}  \sum_{l \in \U} d_l f_l^u \label{32} \\
\mbox{s.t. } A^uf^u = b, f^u \ge 0. \label{33}
\end{gather}
If $VMT^m$ is the first term in \eqref{27}, we can estimate total VMT as
\[VMT \approx VMT^m \pm \frac{1}{2	} [VMT^u_+ - VMT^u_-].\]

\subsection{Case study continued}
\begin{figure}[h!]
\centering
\includegraphics[width=6.5in]{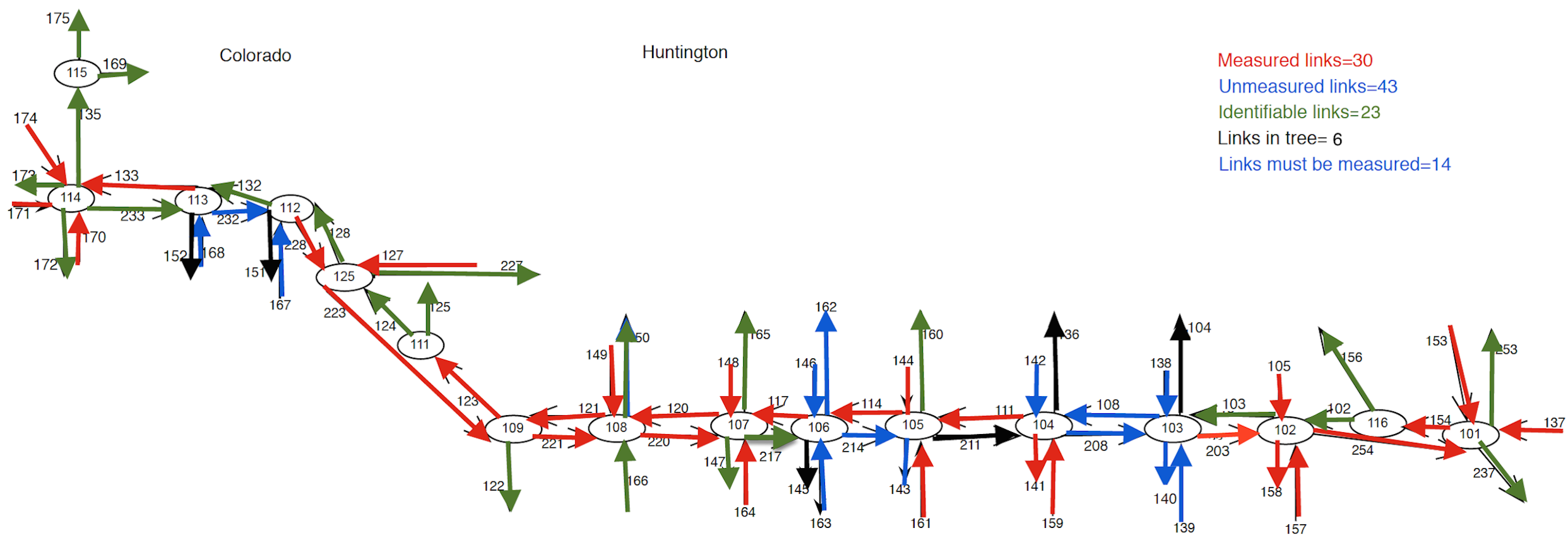}
\caption{Measured and identifiable links, and 14 links that need to be measured.}
\label{fig:VMT}
\end{figure}

The network of Figure \ref{fig-site} has 16 nodes, 73 links, and 106 turn movements (the other turn movements such as U-turns are zero) for
a total of 179 flows.  
The measurements comprise 30 link flows and 57 turn ratios.  Application of the procedures in Theorem \ref{T1} and Corollary \ref{C3} lead to
the identification of the flows on 23 unmeasured links. This leaves 73-30-23 = 20 link flows that cannot be identified from the measurements.  In Figure \ref{fig:VMT} the measured links are colored
red, the unmeasured but identifiable links are colored green, the  unidentifiable links are colored blue or black.  
The graph comprising these unidentifiable links has 7 nodes (including node 0, not shown), so by Corollary \ref{C1} an additional (20 - (7-1)) = 14 link flows must be measured to
completely identify all network flows.  We select any tree in this graph, the (7-1) = 6 black links is one such tree; the remaining 14 blue links must be measured to identify all link flows. 
Ten of the blue links are exit or entry links and 4 are internal links.
Figure \ref{fig:VMT} summarizes these findings.  

If it is not possible to make these 14 additional measurements, we can use \eqref{30}-\eqref{33} to bound the inaccuracy.  We find the maximum
and minimum VMT compatible with all the measurements to be
\begin{equation} \label{34}
VMT^m + VMT^u_+ = 6824 VMT/hour, \; VMT^m + VMT^u_- = 4998 VMT/hour,
\end{equation}
which gives an error bound of $\pm 15\%$.  
We simulate the baseline model for 3 hours and record the sequence $(d_i,t_i)$ of travel distance and travel time for
every vehicle $i$ from the instant it enters the network until it leaves the network.

The simulation gives per hour values of 
VMT =1/3 $\sum_i d_i = 5,613$ veh-miles, which conforms with the bounds in \eqref{34},  
VHT = 1/3 $\sum_i t_i = 894$ veh-hours, and an estimated average speed of
\[v = \mbox{VMT/VHT = 6.28 mph}.\]
On the other hand, we can directly calculate the average of all vehicle trip lengths, travel times, and experienced speeds as
\[\bar{d} = \frac{1}{N} \sum_{i=1}^N  d_i =0.6 \mbox{ miles}, \bar{t} = \frac{1}{N} \sum_{i=1}^N  t_i = 114.3 \mbox{ s}, \bar{v} = \frac{1}{N} \sum_{i=1}^N v_i = 8.95 \mbox{ mph}.\]
Here $v_i = d_i/t_i$ and $N = 28,153$ is the total number of trips that traverse at least one internal link (trips that start at an entry link and immediately exit are omitted).  
Since we find $v < \bar{v} $, this must imply that 
\[\frac{\sum t_iv_i}{\sum t_i} < \frac{\sum v_i}{N},\]
and this inequality suggests that $\{v_i\}$ and $\{t_i\}$ are negatively correlated: longer-duration trips  are made at slower speeds, and indeed calculation gives the correlation $\rho(v_i, t_i) = -0.15$.

\section{A macroscopic queuing model} \label{sec-macro}
\begin{figure}[h!]
\centering
\includegraphics[width=5in]{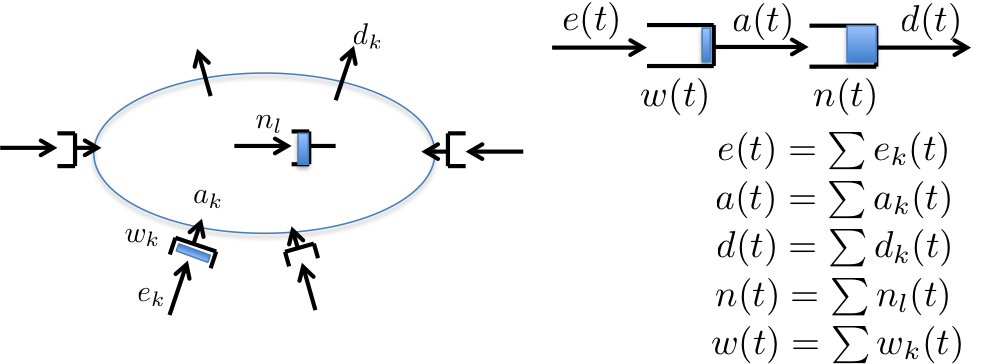}
\caption{Macroscopic queuing model of arterial network, with aggregate quantities $e, a, d, w, n$.}
\label{fig:macroq}
\end{figure}
In the left of Figure \ref{fig:macroq}, the $\{e_k(t)\}$ are \textit{external} arrivals into the entry links, the
$\{a_k(t)\}$ are \textit{internal} arrivals into the network upon leaving the entry link queues, the $\{d_k(t)\}$ are \textit{departures} from the network, the $\{w_k(t)\}$ are the numbers of vehicles queued in the entry links, and the
$\{n_k(t)\}$ are the numbers of vehicles in the internal links, either queued or traveling.  The aggregate quantities $e (t), a(t), d(t), w(t), n(t)$ are processes of the macroscopic queuing system shown in the right of the Figure \ref{fig:macroq}. 
The behavior of the macroscopic queue depends on two factors -- the pattern of demand (volumes, turn ratios) and the intersection signal controllers -- as will be seen in the analysis of two experiments.

\begin{figure}	
	\centering
	\begin{subfigure}[t]{3in}
		\centering
		\includegraphics[height=2.35in]{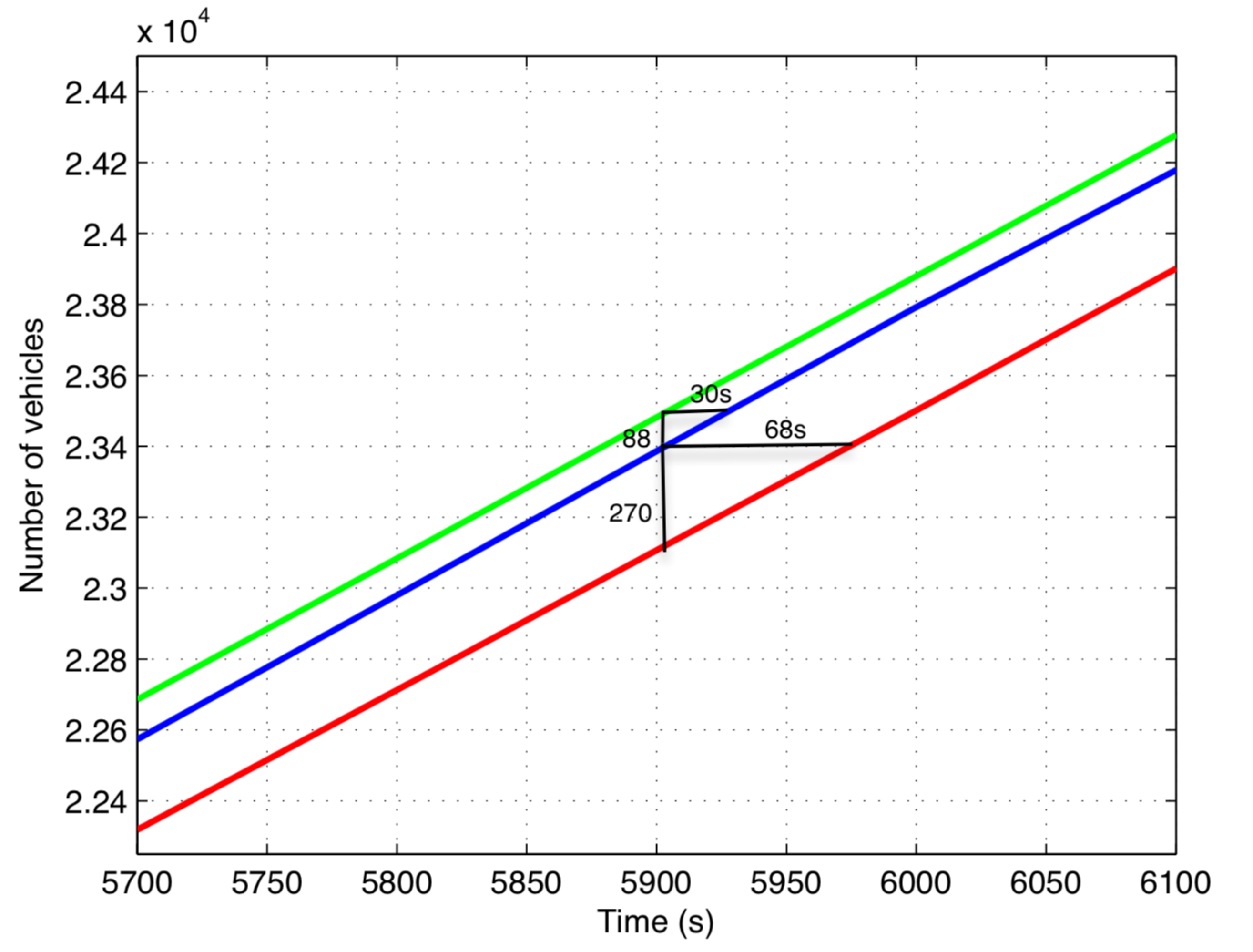}
		\caption{}\label{fig:a}		
	\end{subfigure}
	\quad
	\begin{subfigure}[t]{3in}
		\centering
		\includegraphics[height=2.4in]{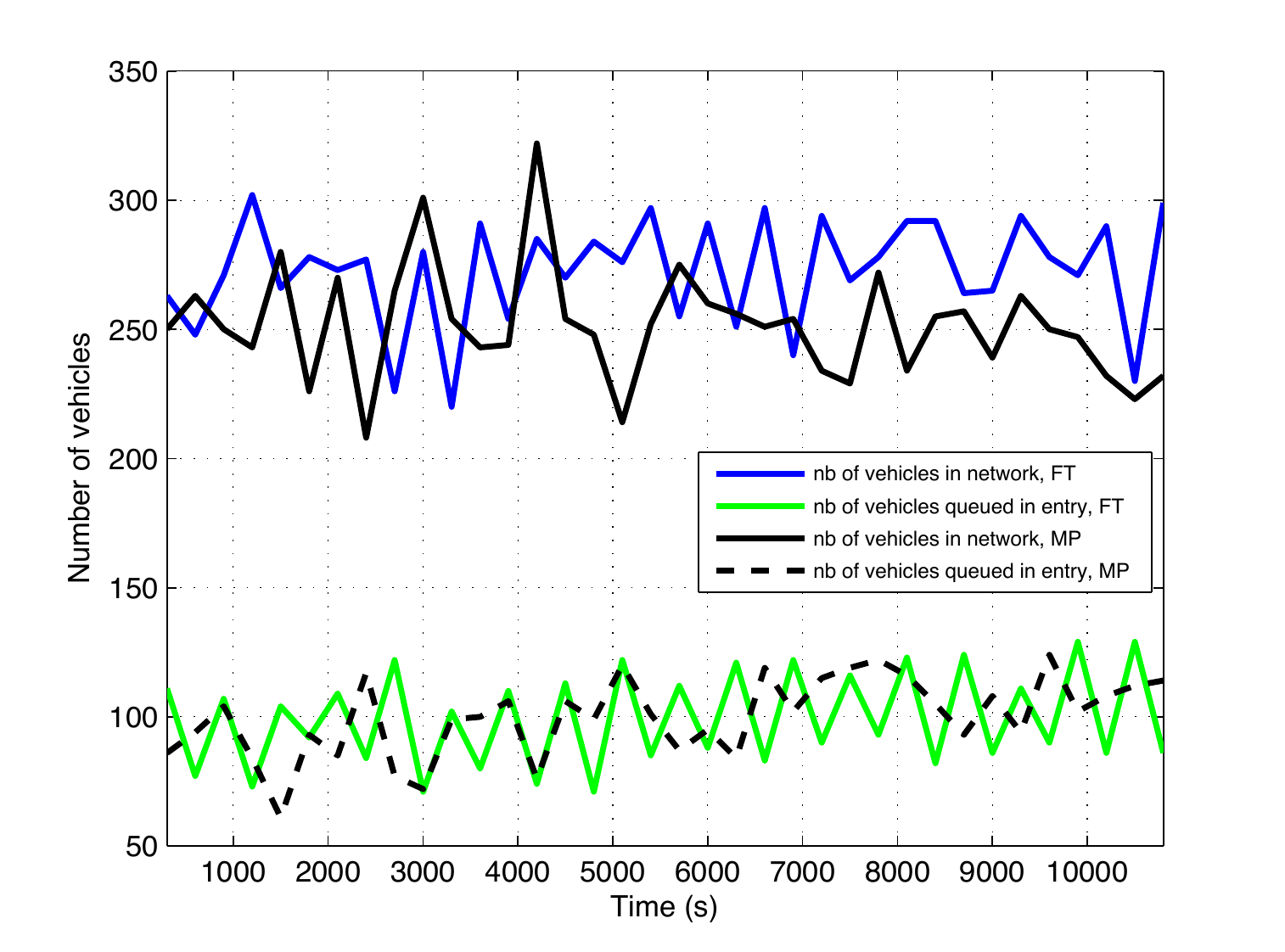}
		\caption{ }\label{fig:b}
	\end{subfigure}
	\caption{(a) Cumulative  external arrivals (green), internal arrivals (blue) and departures  (red) for FT control and $\gamma =1$, (b) number of vehicles,$w(t)$ and $n(t)$ for FT control and MP control.}
\label{fig:12}
\end{figure}

In the first experiment we load the network with the baseline arrivals  $\{e_k (t)\}$, with turn ratios also given by the baseline model.  The total external arrival rate is $ \lambda =  \sum e_k (t) = 14,335$ vph.  With this loading the network is stable, so the internal arrival and departure rates  also equal $\lambda$.   If $E$ is the average time spent in the entry links, $T$ is the average sojourn or trip time inside the network, and $w, n$ are the average number of vehicles waiting in the entry links and inside the network, Little's formula gives
\begin{equation} \label{35}
w+ n = (E +T) \lambda; \quad w = E \lambda;\quad n = T \lambda.
\end{equation}
Figure \ref{fig:12}(a) plots the cumulative external arrivals (green), internal arrivals (blue) and departures (red) during the time interval [5700,6100]s.  The slope of all three curves is $\lambda = 14,335$ vph.  The horizontal (vertical) distance between the green and red curves is $E+T = 98$s ($w+n = 358$ vehicles); and  the horizontal (vertical) distance between the blue and red curves is $T = 68$s ($n = 270$ vehicles), in accordance with \eqref{35}.  We also consider the behavior of the network under max pressure (MP) control.  From 
Figure \ref{fig:12} (b)  we see that the average value of  $n(t)$ for  FT is considerably larger than for MP, as expected.  However, since the network is stable ($w(t)$ and $n(t)$ remain bounded) under both controls, arrivals into and departures from the network are virtually the same.

In the second experiment, we  increase the network loading proportionately in 2-hour steps as
$\{\gamma e_k (t)\}$, with $\gamma$ =1, 1.05, 1.15, 1.3, 1.5, 1.7, 2, 2.2, 2.5.  
Because 2 hours is much longer than the longest travel times (see Figure \ref{fig:TT}), the network will reach a `steady state' if it is stable.  From Figure \ref{fig:13} we can see that under FT, the network is stable for $\gamma = 1.05$ ($t=2$h), whereas under MP, it is stable for $\gamma = 1.7$ ($t = 12$h).  For $\gamma = 1.05$ ($t= 2$h), the network is stable under both FT and MP controls, and the macroscopic queues are similar, like in Figure \ref{fig:12}(b). 

However for $\gamma > 1.05$, the FT controlled network becomes  unstable, some internal links  become saturated,  causing spill back and blocking of upstream links.  Eventually vehicles  queue up at the entry links, the internal arrival rate  drops below the external arrival rate (at $t=2$h),  and the number of vehicles inside the network  increases as it gets congested.  This is seen in Figure \ref{fig:13}(c).  

The MP-controlled network becomes unstable for $\gamma > 1.7$ ($t=12$ h):
internal queues become saturated, vehicles queue up at the entry links and the departure rate drops (Fig \ref{fig:13}(b),(c)). 

Figure \ref{fig:13} suggests that we may define the \textit{network capacity} as the maximum value of $\g \times 14,335$ for which the network is stable, which turns out to be 15,051 vph ($\gamma =1.05$) for FT and 24,369 vph ($\gamma =1.7$) for MP.  
Thus network capacity depends on the intersection control.  It also depends on the pattern of demand.\footnote{Although not shown here, a demand pattern different from the baseline, leads to different network capacity.}
 
Once the external arrival rate exceeds capacity, the network queue $n(t)$ rises very sharply, as seen in Figure \ref{fig:13}(c).  This rapid increase resembles a `phase transition'.  Detailed analysis of the simulations suggests the following pattern.  When network loading exceeds capacity, some internal links get saturated, blocking upstream links.  Congestion spreads and in a short time a significant portion of the network links becomes saturated, causing a sharp increase in  the number of vehicles in the network.  We might say that as external demand exceeds network capacity, the network rapidly enters `gridlock'.

\begin{figure}[h!]
\centering
\includegraphics[width=5in]{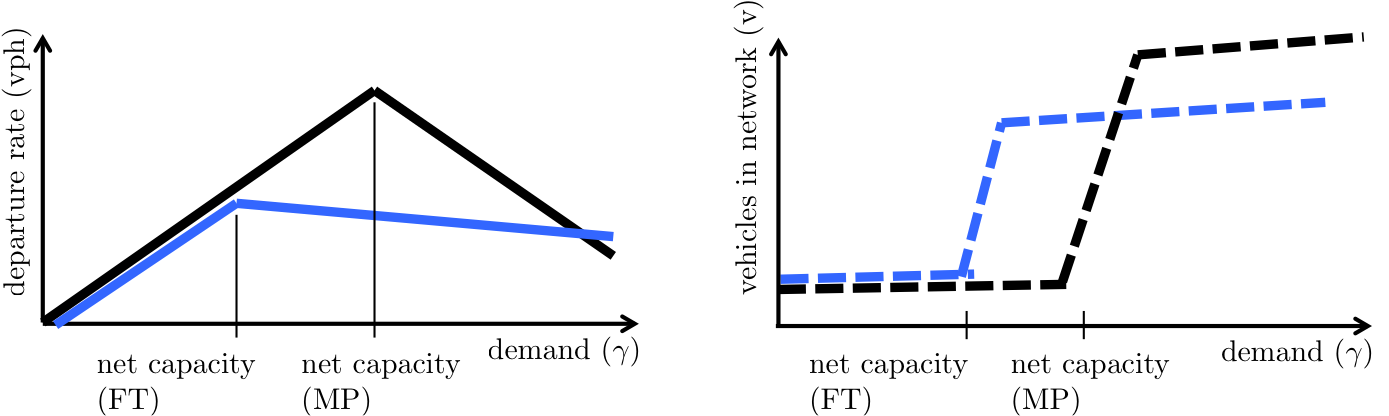}
\caption{Macroscopic queuing diagram (MQD) of arterial network under FT and MP control: departure rate (left) and number of vehicles in the network (right) as
demand $\gamma$ is increased.}
\label{fig:mqm}
\end{figure}

In summary, the behavior of the macroscopic queue model (MQM) can be characterized in terms of network capacity: if the external demand is below capacity, the network is stable,  queues remains bounded and the rate of arrivals (demand) equals the rate of departures.  When demand exceeds the network capacity, the network becomes saturated, and the rate of departures drops below the demand.   Thus if we were to plot aggregate rate of departures ($d$) as a function of the aggregate rate of external arrivals $e$, $d = f(e)$, the graph of $f$ will have slope of 1 until $e$ reaches capacity, after which the slope drops below 1 and vehicles begin to accumulate in the network.  Moreover, the number of vehicles in the network will be relatively constant until demand reaches capacity after
which it grows rapidly.  This suggests the stylized macroscopic queuing diagram (MQD) of Figure \ref{fig:mqm}.

\subsection{Comparison with macroscopic fundamental diagram (MFD)}
It seems appropriate to  compare the macroscopic queue model (MQM) and the macroscopic fundamental diagram (MFD) of \cite{mfd}.   MFD is an empirical device that captures the relationship between the aggregate flow and density within a network, using measurements of flow $f_i(t)$ and occupancy $o_i(t)$ for a sample of links $i$.  The aggregate flow and density are constructed as
\begin{equation}\label{36}
f (t)= \frac{\sum_i l_i f_i(t)}{\sum_i l_i}, \quad o (t)= \frac{\sum_i l_i o_i (t)}{\sum_i l_i},
\end{equation}
in which $l_i$ is the length of link $i$ multiplied by the number of lanes.  The scatter plot of pairs $(f(t), o(t))$, $t = 1, \cdots, T$ has a parabolic shape, similar to what one gets for a single link.  This plot is called the macroscopic fundamental diagram (MFD).  

MFD and MQM are  related.  MFD's  aggregate occupancy measure  is related to the number of vehicles in MQM: indeed occupancy as measured by detectors is a proxy for spatial density, which is simply number of vehicles in a link divided by its length.  MFD's  flow measure is related to the aggregate  departure rate of MQM, for as  \cite{mfd}  argue there is a `linear relation' between departure rate and network flow.  The notion of network capacity in both MFD and MQM are similar: MFD defines it as the maximum network flow, and MQM defines it as maximum departure rate.  

One difference between MFD and MQM is that the latter makes explicit its dependence on both the pattern of demand and signal control.  This dependence is only implicit in MFD: for example, MFD for different time periods of the day will be different since the pattern of traffic is different.
Another difference concerns network behavior when demand exceeds capacity.  Since MFD is an equilibrium notion, it cannot describe the phase transition that accompanies instability caused by excess demand, whereas MQM is a dynamic model that does portray what happens in excess demand conditions.

Lastly, both MFD and MQM suggest the value of `perimeter control' to prevent 
 congestion inside the network (\cite{geroliminis-perimeter}).  Indeed it is clear from Figures \ref{fig:13} and \ref{fig:mqm} that if external arrivals are restricted so that they do not  exceed capacity, the  network will not get congested.  As Figure \ref{fig:13}(b) suggests, 
appropriate perimeter control will lead to an accumulation of vehicles at entry links in case of excess demand while protecting the internal network from congestion.  Implementing a perimeter control requires having a signal that indicates when 
network capacity is approached.  It seems difficult to directly measure the aggregate arrival rate since an urban network will have
many entry links.  If the number of vehicles, $n(t)$, inside the network could be measured, a rapid increase
in $n(t)$ would provide such a signal as seen in Figure \ref{fig:13}(c).  However, measuring $n(t)$ seems impracticable.  One
proxy measure is the travel time $T$ experienced by vehicles going through the network, for according to \eqref{35} $T$ is
proportional to the aggregate arrival rate.

\section{Conclusion} \label{sec-conc}
Microsimulation studies of arterial networks must compromise between adopting a model that is rich in behavioral detail and working with field measurements that are  incomplete and sometimes inconsistent.  Using a behaviorally rich model  may entail abandoning rigorous calibration or validation.  This paper proposes an alternative approach based on the PointQ simulation model.  PointQ is a mesoscopic-microscopic model: it maintains the identity of individual vehicles, but ignores inter-vehicle interaction due to driver car-following behavior.  PointQ's advantage is that it is easy to calibrate using commonly available field data.  The approach uses conservation of flow identities to impute  unmeasured flows, and it provides bounds in terms of a VMT gap on the inaccuracy due to flows that cannot be imputed and gives minimal sets of additional flow measurements to impute all flows.  Most of the effort in using PointQ is devoted to evaluating performance  (such as travel times and intersection  performance) that can be compared with other field measurements or with the intuition that engineers operating the network have.
The rest of the effort is devoted to using the calibrated PointQ model to conduct experiments for which the model is constructed.

The approach is illustrated by a small arterial network adjacent to the I-210 freeway in Los Angeles.  The network has 16 intersections, 73 links and 106 turn movements, for a total of 179 flows.\footnote{The network studied here is part of a  network with 450 intersection, so it is 30 times larger.  The calibration algorithms and the PointQ simulator itself are expected to scale linearly to the larger network.}   The measurements comprise flows on only 30 links and 57 turn ratios.  The calibration and imputation procedure identifies all except  22 flows and determines that 14 additional measurements are needed to identify all flows.  Without these measurements there is an error bound of $\pm 15\%$  on VMT that is accounted by the model.  Several experiments are conducted.  Two of these determine how much additional freeway traffic can be diverted on to the arterial network with or without a change in timing plans.  The third experiment estimates the savings in intersection delay, travel time, and network capacity obtained by replacing the fixed time controllers by max pressure controllers.  The fourth experiment demonstrates that it is a poor idea to model traffic as a single-commodity flow when you know that the traffic follows two different routing disciplines.  

A macroscopic queuing model (MQM) is proposed to relate
aggregates of external, internal and departure flows, together with the total number of vehicles in the network.  MQM is compared with the macroscopic fundamental diagram (MFD). Both MQM and MFD lead to the notion of network capacity which, in turn, is useful in constructing a perimeter control to prevent network congestion.

\setcounter{equation}{0}
\appendix
\appendixpageoff
\section*{Appendix} 
\begin{appendices}
\renewcommand{\theequation}{\Alph{section}.\arabic{equation}}

The following table show the calibration results obtained by solving \eqref{10}-\eqref{14}. 

\includepdf[pages={1,2}]{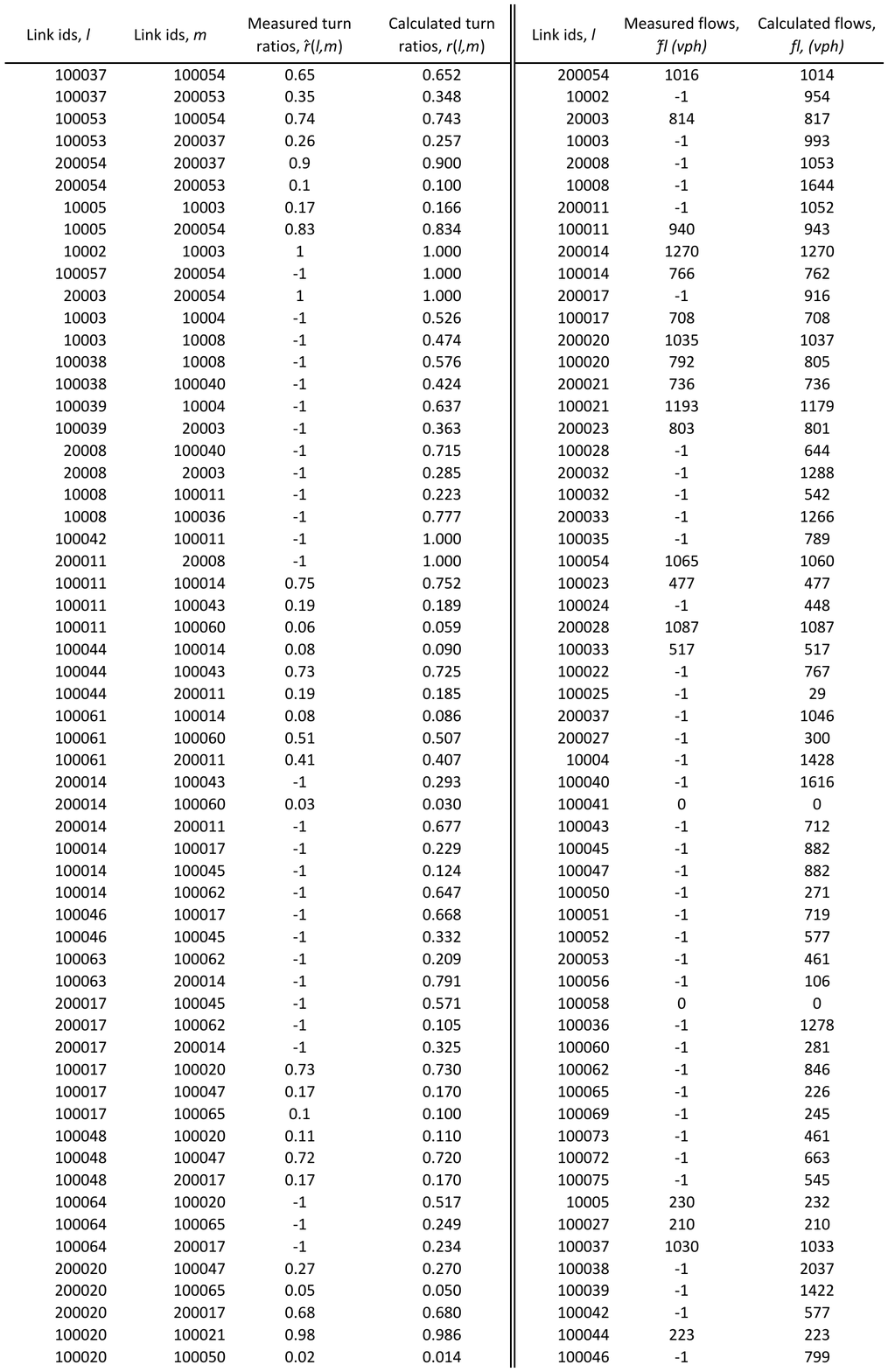}

\end{appendices}
\bibliographystyle{plainnat}

\bibliography{traffic,traffic1,traffic2}

\providecommand{\url}[1]{#1}  \providecommand{\url}[1]{#1}
  \providecommand{\url}[1]{#1}
\begin{thebibliography}{12}
\providecommand{\natexlab}[1]{#1}
\providecommand{\url}[1]{\texttt{#1}}
\expandafter\ifx\csname urlstyle\endcsname\relax
  \providecommand{\doi}[1]{doi: #1}\else
  \providecommand{\doi}{doi: \begingroup \urlstyle{rm}\Url}\fi

\bibitem[Aboudolas et~al.(2009)Aboudolas, Papageorgiou, and
  Kosmatopoulos]{aboudolas09A}
K.~Aboudolas, M.~Papageorgiou, and E.~Kosmatopoulos.
\newblock Store-and-forward based methods for the signal control problem in
  large-scale congested urban road networks.
\newblock \emph{Transportation Research, Part C}, 17:\penalty0 163--174, 2009.

\bibitem[Geroliminis and Daganzo(2008)]{mfd}
N.~Geroliminis and C.~F. Daganzo.
\newblock Existence of urban-scale macroscopic fundamental diagrams: Some
  experimental findings.
\newblock \emph{Transportation Research Part B}, 42\penalty0 (9):\penalty0
  759--770, 2008.

\bibitem[Geroliminis et~al.(2013)Geroliminis, Haddad, and
  Ramezani]{geroliminis-perimeter}
N.~Geroliminis, J.~Haddad, and M.~Ramezani.
\newblock Optimal perimeter control for two urban regions with macroscopic
  fundamental diagrams: A model predictive approach.
\newblock \emph{IEEE Transactions on Intelligent Transportation Systems,},
  14\penalty0 (1):\penalty0 348--359, 2013.

\bibitem[Jayakrishnan et~al.(1994)Jayakrishnan, Mahmassani, and
  Lin]{jayakrishnan1994}
R.~Jayakrishnan, H.S. Mahmassani, and {T.-Y.} Lin.
\newblock An evaluation tool for advanced traffic information and management
  systems in urban networks.
\newblock \emph{Transportation Research C}, 2\penalty0 (3):\penalty0 129--147,
  1994.

\bibitem[Kouvelas et~al.(2014)Kouvelas, Lioris, Fayazi, and Varaiya]{liorisTRR}
T.~Kouvelas, J.~Lioris, S.A. Fayazi, and P.~Varaiya.
\newblock Maximum pressure controller for stabilizing queues in signalized
  arterial networks.
\newblock \emph{Transportation Research Record}, 2421:\penalty0 133--141, 2014.

\bibitem[Lioris et~al.(2014)Lioris, Kurzhanski, Triantafyllos, and
  Varaiya]{wodes}
J.~Lioris, A.A. Kurzhanski, D.~Triantafyllos, and P.~Varaiya.
\newblock Control experiments for a network of signalized intersections using
  the {.Q} simulator.
\newblock {12th IFAC - IEEE International Workshop on Discrete Event
  Systems(WODES)}, May 2014.

\bibitem[Lo(2001)]{lo2001}
H.K. Lo.
\newblock A cell-based traffic control formulation: Strategies and benefits of
  dynamic timing plans.
\newblock \emph{Transportation Science}, 35\penalty0 (2):\penalty0 148--164,
  2001.

\bibitem[Muralidharan et~al.(2014{\natexlab{a}})Muralidharan, Flores, and
  Varaiya]{ITSC2014}
A.~Muralidharan, C.~Flores, and P.~Varaiya.
\newblock High-resolution sensing of urban traffic.
\newblock In \emph{IEEE 17th International Conference on Intelligent
  Transportation Systems (ITSC)}, pages 780--785, Qingdao, China, October 8-11
  2014{\natexlab{a}}.

\bibitem[Muralidharan et~al.(2014{\natexlab{b}})Muralidharan, Pedarsani, and
  Varaiya]{FTControl}
A.~Muralidharan, R.~Pedarsani, and P.~Varaiya.
\newblock Analysis of fixed-time control, Nov 2014{\natexlab{b}}.
\newblock URL \url{http://arxiv-web3.library.cornell.edu/abs/1408.4229v2}.
\newblock To appear in Transportation Research B.

\bibitem[Park and Schneeberger(2004)]{bpark}
B.~Park and J.~D. Schneeberger.
\newblock Microscopic simulation model calibration and validation.
\newblock \emph{Transportation Research Record}, 1856:\penalty0 185--192, 2004.

\bibitem[{Trasportation Research Board}(2010)]{HCM}
{Trasportation Research Board}.
\newblock \emph{{Highway Capacity Manual}}.
\newblock Washington DC, 2010.

\bibitem[Varaiya(2013)]{MPtrc}
P.~Varaiya.
\newblock Max pressure control of a network of signalized intersections.
\newblock \emph{Transportation Research, Part C}, 36:\penalty0 177--195, 2013.

\end{thebibliography}

\end{document}